\documentclass[twocolumn,secnumarabic,amssymb, nobibnotes, aps, prb, superscriptaddress]{revtex4-2}
\usepackage{lipsum}
\usepackage{titlesec}
\titlespacing\section{0pt}{12pt plus 4pt minus 4pt}{1pt plus 20pt minus 2pt}
\usepackage{xcolor}
\usepackage[colorlinks, allcolors=blue]{hyperref}
\usepackage{tabularx}
\usepackage{amsmath}
\usepackage{comment}
\usepackage{chemformula}
\usepackage{afterpage}
\usepackage{placeins}
\usepackage{graphicx}
\usepackage{float}
\usepackage{booktabs}
\usepackage{multirow}
\usepackage{array}
\usepackage{setspace}
\graphicspath{{Figs/}}
\usepackage{siunitx}
\usepackage{hhline}
\usepackage{xfrac}
\usepackage{float,graphicx}
\usepackage{mathtools}
\usepackage{listings}
\usepackage{amssymb}
\usepackage{titlesec}
\usepackage{amsfonts}
\usepackage[version=4]{mhchem}

\usepackage{epstopdf}
\catcode`@11
\def\seceqaa{\@addtoreset{equation}{section}
\def\theequation{A\arabic{equation}}}
\def\seceqbb{\@addtoreset{equation}{section}
\def\theequation{B\arabic{equation}}}
\def\seceqcc{\@addtoreset{equation}{section}
\def\theequation{C\arabic{equation}}}
\def\seceqdd{\@addtoreset{equation}{section}
\def\theequation{D\arabic{equation}}}
\def\seceqee{\@addtoreset{equation}{section}
\def\theequation{E\arabic{equation}}}
\def\seceqff{\@addtoreset{equation}{section}
\def\theequation{F\arabic{equation}}}
\def\seceqgg{\@addtoreset{equation}{section}
\def\theequation{G\arabic{equation}}}
\def\seceqhh{\@addtoreset{equation}{section}
\def\theequation{H\arabic{equation}}}
\catcode`@11

\begin{document}

\title{Giant anisotropic anomalous Hall effect in antiferromagnetic topological metal NdGaSi}

\author{Anyesh Saraswati} 
\altaffiliation{These authors contributed equally to this work}
\affiliation{S. N. Bose National Centre for Basic Sciences, JD Block, Sector III, Salt Lake, Kolkata 700106, India}

\author{Sudipta Chatterjee} 
\altaffiliation{These authors contributed equally to this work}
\affiliation{S. N. Bose National Centre for Basic Sciences, JD Block, Sector III, Salt Lake, Kolkata 700106, India}

\author{Nitesh Kumar}\email{nitesh.kumar@bose.res.in}
\affiliation{S. N. Bose National Centre for Basic Sciences, JD Block, Sector III, Salt Lake, Kolkata 700106, India}


\begin{abstract}
The interplay between magnetism and strong electron correlation in magnetic materials discerns a variety of intriguing topological features. Here, we report a systematic investigation of the magnetic, thermodynamic, and electrical transport properties in NdGaSi single crystals. The magnetic measurements reveal a magnetic ordering below \textit{T}$_N$ ($\sim$ 11 K), with spins aligning antiferromagnetically in-plane, and it orders ferromagnetically (FM) out-of-plane. The longitudinal resistivity data and heat capacity exhibit a significant anomaly as a consequence of the magnetic ordering at \textit{T}$_N$. The magnetoresistance study shows significantly different behavior when measured along either direction, resulting from the complex nature of the magnetic structure, stemming from complete saturation of moments in one direction and subsequent spin flop transitions in the other. Remarkably, we have also noticed an unusual anisotropic anomalous Hall response. We have observed a giant anomalous Hall conductivity (AHC) of $\sim$ 1730 $\Omega^{-1}$ cm$^{-1}$ and  $\sim$ 490 $\Omega^{-1}$ cm$^{-1}$ at 2 K, with $B \parallel $ [001] and $B \parallel $ [100], respectively. Our scaling analysis of AHC reveals that the anomalous Hall effect in the studied compound is dominated by the Berry phase-driven intrinsic mechanism. These astonishing findings in NdGaSi open up new possibilities for antiferromagnetic spintronics in rare-earth-based intermetallic compounds.

\end{abstract}

\maketitle
\section{INTRODUCTION}
\vspace{3mm}
In correlated electronic systems, the hybridization of 4\textit{f} and conduction electrons gives rise to unique physical characteristics such as density wave, superconductivity, heavy-fermionic behavior, and complex magnetism. \cite{steglich1979superconductivity,mathur1998magnetically,sacchetti2007pressure,brouet2008angle,gaudet2021weyl,cheng2024tunable}. Consequently, rare-earth-based inter-metallic compounds offer an ideal platform for investigating a wide range of these novel physical properties. Recently, the family of rare-earth-based ternary compounds \textit{R}Al\textit{X} (\textit{R} = La–Nd and Sm; \textit{X} = Si and Ge) has gathered a lot of interest due to their potential Weyl semimetallic state \cite{chang2018magnetic,suzuki2019singular,yang2021noncollinear,puphal2020topological,lyu2020nonsaturating,yao2023large,yamada2024nernst}. The spontaneous magnetic ordering and electronic band topology of these materials are primarily determined by the rare earth ions. Moreover, their magneto-transport properties are also quite different from each other. To illustrate, CeAlSi displays an unusual anisotropic anomalous Hall response along the easy and hard magnetic axes \cite{yang2021noncollinear}, while SmAlSi exhibits a large topological Hall response accompanied by a spiral magnetism mediated by Weyl fermions \cite{yao2023large}. Similarly, magnetic Weyl semimetal NdAlSi shows anomalous angular magnetoresistance and an unusual quantum oscillation \cite{wang2022ndalsi,wang2023quantum}, while PrAlSi demonstrates a large nonsaturating magnetoresistance and anomalous Hall conductivity \cite{lyu2020nonsaturating}. Furthermore, it has also been observed that NdAlGe indicates a long-wavelength helical magnetism \cite{gaudet2021weyl}, whereas CeAlGe reveals that the topological characteristics of these systems are strongly dependent on their chemical composition \cite{piva2023topological}.

Lately, another related family of these ternary rare earth compounds \textit{R}Ga\textit{X} (\textit{R} = rare-earth elements, \textit{X} = Si and Ge) have also attracted much attention due to their fascinating magnetic and transport properties \cite{ram2023magnetic,gong2024anomalous,zhang2024magnetism}. For instance, CeGaGe and PrGaGe display a complex magnetic structure along with a strong magnetocrystalline anisotropy \cite{ram2023magnetic}. On the other hand, CeGaSi emerges as a Kondo system with heavy fermionic behavior \cite{zhang2024magnetism}; additionally, it shows the absence of anomalous Hall in one direction and skew-scattering-dominated anomalous Hall effect in the other \cite{gong2024anomalous}. These compelling characteristics of the \textit{R}Ga\textit{X} family encourage us to investigate the other members of the family and study the anomalous magnetic and transport properties.

The anomalous Hall effect (AHE) in antiferromagnets has been a topic of immense interest lately \cite{nagaosa2010anomalous,kubler2014non,aheafm2022libor}. Antiferromagnets with zero net magnetic moments generally do not exhibit AHE. However, AFMs such as Mn$_3$Sn and Mn$_3$Ge with very small net spontaneous moment exhibit large AHE because of the Berry curvature effect \cite{nakatsuji2015large,nayak2016large}. Subsequently, a large value of AHE was also observed in AFM frustrated kagom{\'e} lattice, where the effect was mediated by the formation of Weyl points in the band structure due to the application of a magnetic field which, in turn, dramatically enhances the Berry curvature due to the breaking of time-reversal symmetry (TRS). \cite{zhao2020ice,zhao2024ahe,roychowdhury2024enhancement}. Our system, NdGaSi, is an excellent candidate to study the AHE in an AFM system, as we observe a sharp magnetic transition leading to saturation of moments on the application of a very small magnetic field, breaking the TRS, a prerequisite to finite net Berry curvature effect. Nevertheless, the experimental investigation of topology-driven anomalous transport properties in this novel ternary \textit{R}Ga\textit{X} family is still lacking in the literature.

In this present study, we report the magnetic, specific heat, and magneto-transport properties of single crystalline NdGaSi. The magnetic and electrical resistivity data indicate the onset of antiferromagnetic ordering below 11 K, which is further corroborated by a significant anomaly in specific heat. The magnetotransport measurement illustrates quite different behavior in both directions due to the complex magnetic structure, which includes complete spin polarization in one direction and spin flop transitions in the other. Interestingly, we observe an anisotropic and giant anomalous Hall response in NdGaSi. The anomalous Hall conductivity (AHC) reaches up to $\sim$ 1730 $\Omega^{-1}$ cm$^{-1}$ and  $\sim$ 490 $\Omega^{-1}$ cm$^{-1}$ at 2 K, with $B \parallel $ [001] and $B \parallel $ [100], respectively. The intrinsic AHC of 1166 $\Omega^{-1}$ cm$^{-1}$ in $B \parallel $ [001] orientation is significantly larger than that of the previously reported AFMs and comparable to FMs. 
\vspace{3mm}
\section{EXPERIMENTAL METHODS}
\vspace{3mm}

High-quality single crystals of NdGaSi were synthesized using a gallium self-flux method. Pieces of neodymium (Nd), gallium (Ga), and silicon (Si) were mixed in a molar stoichiometric ratio of 1:10:1 and put in an alumina crucible. The crucible was sealed in a quartz tube, with quartz wool as a filter, under partial argon pressure and then put in a muffle furnace. The ampoule was heated to 1100$ ^\circ$ C at a rate of 100$ ^\circ$ C/h, where it was dwelt for 12 h. The reaction was cooled to 600$ ^\circ$ C at a rate of 3$ ^\circ$ C/h and the excess gallium flux was decanted using a centrifuge. Shiny, rectangular, plate-like crystals of NdGaSi were obtained. Single crystals of a zero 4\textit{f} local moment counterpart LaGaSi, were also synthesized for specific heat analysis using the gallium self-flux method \cite{gong2024anomalous}. X-ray diffraction (XRD) was carried out on single crystals and powdered single crystals using an x-ray diffractometer (Smart lab, Rigaku) equipped with a 9 kW Cu K$_{\alpha}$ radiation. The obtained XRD data was further refined using the FULLPROF software. To analyze the elemental composition of the obtained crystals, we performed the energy dispersive x-ray spectroscopy (EDX) on a field emission scanning electron microscope (Quanta 250 FEG) equipped with an element silicon drift detector (SDD). Magnetic measurements were performed using the vibrating sample magnetometer (VSM) equipped with a physical properties measurement system (PPMS, Dynacool, Quantum Design). The specific heat measurements were carried out by the PPMS using the conventional thermal-relaxation method. The electrical transport measurements under a magnetic field were performed using the electrical transport option (ETO) of the PPMS. A symmetrization method was employed to eliminate the erroneous transverse contribution, as defined by the formula $\rho_{xx}$(H) = [$\rho_{xx}$(+H) + $\rho_{xx}$(-H)]/2. An antisymmetrization approach was implemented to remove the erroneous longitudinal contribution in the case of Hall resistivity $\rho_{yx}$(H), where $\rho_{yx}$(H) = [$\rho_{yx}$(+H) - $\rho_{yx}$(-H)]/2.

\begin{figure}[t]
\centering
\includegraphics[width=0.48\textwidth]{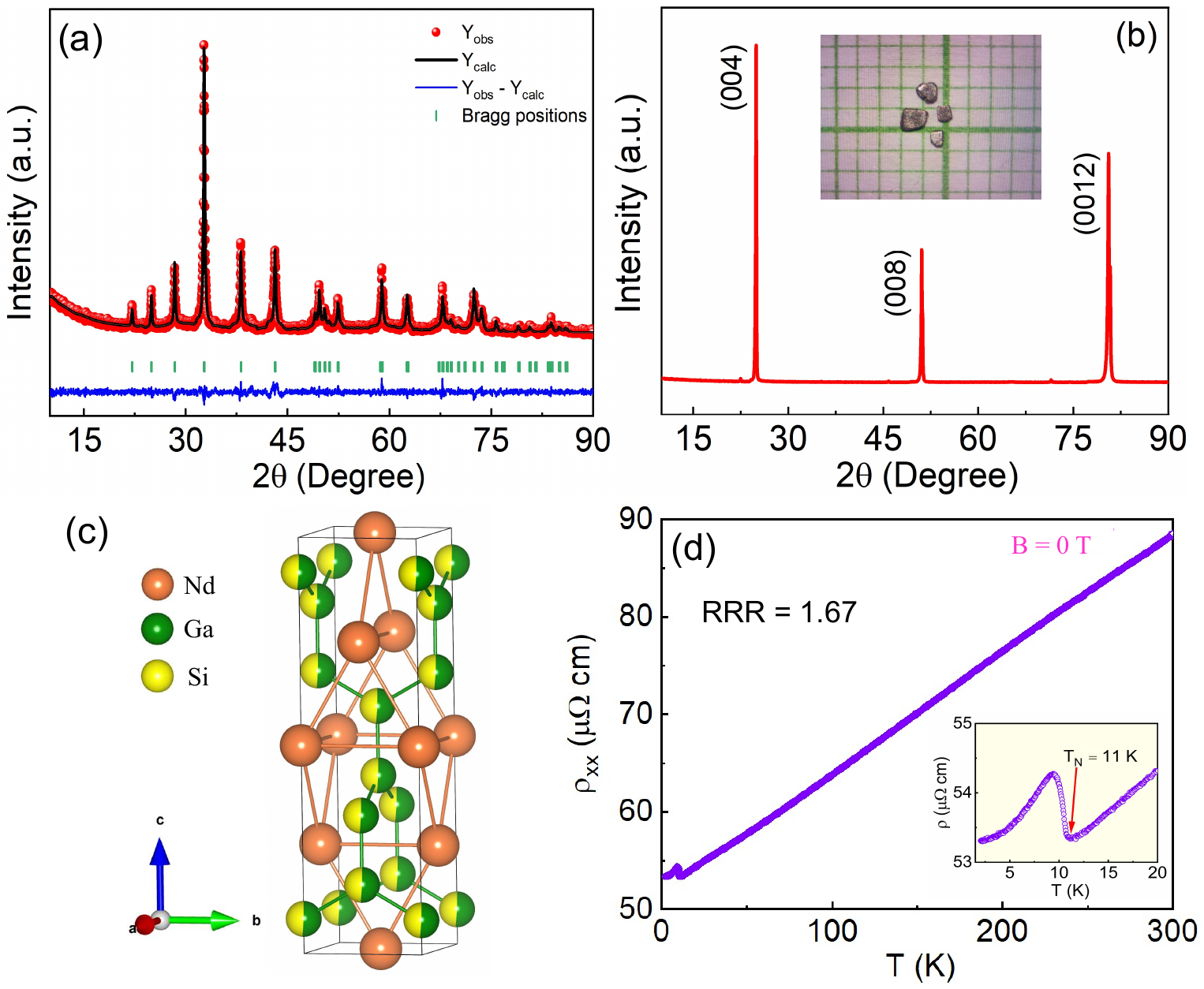}
\caption{(a) Powder XRD pattern of crushed NdGaSi crystals at room temperature. The blue line represents the difference between observed (red) and calculated (black) intensities obtained from Rietveld refinement. The olive vertical lines show the Bragg positions.  (b) Single crystal x-ray diffraction (XRD) data of the (00\textit{l}) plane. Inset shows the picture of the polished NdGaSi crystals (c) Crystal structure of NdGaSi with space group \textit{I}4$_1$\textit{/amd}. (d) Temperature dependence of the longitudinal resistivity ($\rho_{xx}$). The inset shows the magnified view at low temperatures.}
\label{fig1}
\end{figure}

\section{RESULTS AND DISCUSSIONS}
\vspace{3mm}
\subsection{Crystal structure and longitudinal resistivity}
We have carried out EDS spectroscopy to obtain the chemical composition of single crystals. It reveals a near-perfect stoichiometry of Nd:Ga:Si = 1:1.01:0.99 (see Sec.  1 of SI \cite{supply}). Figure \ref{fig1}(a) shows the powder XRD pattern for NdGaSi at room temperature, with no other impurity peaks present. A Rietveld refinement of the powder XRD data supported by single crystal XRD structural analysis indicates a centrosymmetric tetragonal structure with space group \textit{I}4$_1$/\textit{amd}. The obtained lattice parameters are \textit{a} = 4.193  Å  and \textit{c} = 14.285 Å. The room temperature XRD pattern obtained on a single crystal of NdGaSi is depicted in Fig. \ref{fig1}(b). As shown in Fig. \ref{fig1}(c), in the $\alpha$-ThSi$_2$-type centrosymmetric crystal structure, non-rare earth sites are equally preferred by Ga and Si.  In the case of ordering between Ga and Si atoms, the structure will transform into a non-centrosymmetric LaPtSi-type (space group \textit{I}4$_1$\textit{md}) which we do not observe. The crystal was placed with its flat surface parallel to the plane of the sample holder. The observation of only (00\textit{l}) reflections indicates that the c-axis aligns normal with the flat surface.  

Figure \ref{fig1}(d) displays the temperature-dependent longitudinal resistivity $\rho_{xx}$ measured with current \textit{I} along the\textit{ b}-axis ([010]) of NdGaSi single crystal from 2 to 300 K. It exhibits a typical metallic behavior down to $\sim$ 2 K, with a notable anomaly at around 11 K, which is consistent with the magnetic transition (see Fig. \ref{fig2}(a)). A gradual increase in $\rho_{xx}$ below the transition temperature till 9.5 K and a subsequent drop till 2 K. This anomaly may be caused either by enhanced carrier scattering due to spin fluctuations at the onset of transition or because of a magnetic super-zone gap opening in the Fermi surface due to incongruency in the periodicity of the magnetic unit cell to the structural unit cell periodicity. \cite{das2012anisotropic,gupta}. We argue that the anomaly in our resistivity data is caused due to the superzone effect, as the spin fluctuations generally take place just above the magnetic transition, whereas, superzone boundaries in the Brillouin zone form below the ordering temperature, as seen in our case. Compared to conventional metals, NdGaSi has a relatively small residual resistivity ratio (RRR) of $\sim$ 1.66. A value like this for RRR could be the result of site disorder between Si and Ga atoms. Similar RRR has been reported in the \textit{R}Al\textit{X} family (where, \textit{R} = La-Nd, and \textit{X} = Ge, Si) for site disorder between Al and Ge/Si atoms \cite{hodovanets2018single,yang2021noncollinear}.

\subsection{Magnetic properties}

\begin{figure}[t]
\centering
\includegraphics[width=0.47\textwidth]{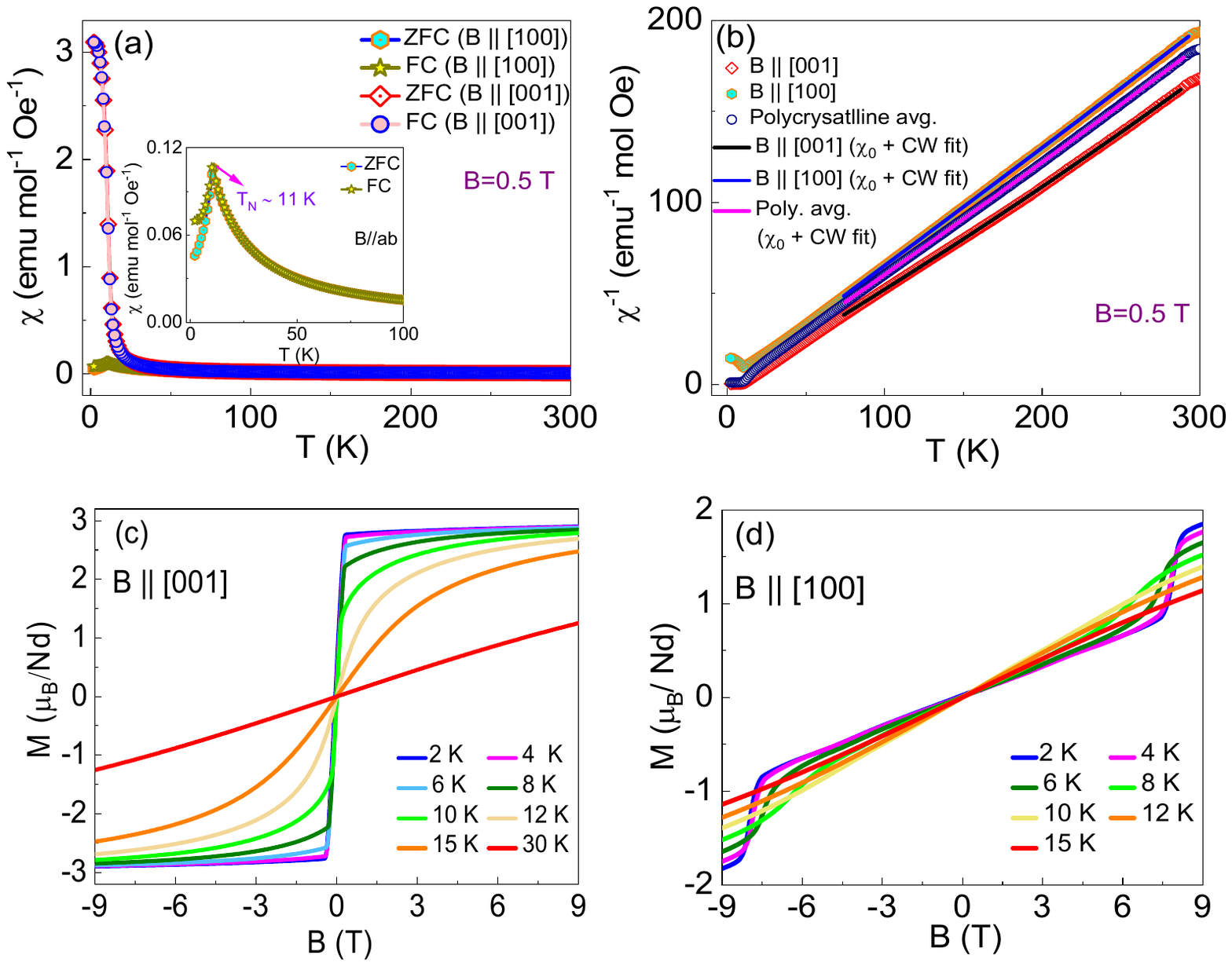}
\caption{(a) Temperature dependence of magnetic susceptibility ($\chi$) with $B \parallel $ [100] and $B \parallel $ [001] for ZFC and FC at B = 0.5 T. (b) Temperature-dependent inverse magnetic susceptibility ($\chi^{-1}$). The solid lines represent the fitting with the modified Curie-Weiss (CW) law. (c) and (d) Field dependence of magnetization M at various temperatures for $B \parallel $ [001] and $B \parallel $ [100], respectively.}
\label{fig2}
\end{figure}

The temperature-dependent magnetic susceptibility curves $\chi$(\textit{T}) for the field-cooled (FC) and zero-field-cooled (ZFC) conditions are displayed in Fig. \ref{fig2}(a), where the magnetic field is applied along [100] and [001] axes. Interestingly, the measured $\chi$(\textit{T}) at $B \parallel $ [001]  is found to be 44 times larger than that measured along  $B \parallel $ [100] at 2 K and 0.5 T, suggesting a strong magneto-crystalline anisotropy, also seen in the case of some other members of the isostructural \textit{R}Al\textit{X} family \cite{yang2021noncollinear,lyu2020nonsaturating,wang2022ndalsi} and slightly higher than the cerium counterpart CeGaSi \cite{gong2024anomalous}. The temperature-dependent FC and ZFC susceptibility curves for $B \parallel $ [100] reveal a maximum at around 11 K, after which they begin to dip downward, suggesting that NdGaSi orders antiferromagnetically along [100], with $T_N$ = 11 K as the N\'{e}el temperature.  Conversely, [001] exhibits an increase in susceptibility when $T_N$ is reached, which suggests ferromagnetic ordering. This could be attributed to a competition between ferromagnetic (FM) and antiferromagnetic (AFM) interactions in the system. The temperature-dependent inverse susceptibility ($\chi^{-1}$) curves are given in Fig. \ref{fig2}(b), and the Curie-Weiss temperatures were determined in both directions by using the modified Curie-Weiss equation to fit the $\chi^{-1}$ vs. \textit{T} curves in the paramagnetic (PM) region between 75 K and 300 K:

\begin{equation}
\chi (T) = \chi_0 + \frac{C}{T - \theta_{W}}
\end{equation}

$\chi_0$ is the temperature-independent magnetic susceptibility, while \textit{C} and $\theta_{W}$ are the Curie constant and paramagnetic Curie temperature, respectively. The fitting of Eq. (1) yields a negative value of $\theta_{W}$ $\sim$ -2.26 K for $B \parallel $ [100] and 0.43 K for $B \parallel $ [001], respectively. The poly-crystalline average was calculated as $\chi_{avg} = \frac{2\chi_{a}+\chi_{c}}{3}$ and fitted with the above equation, where $\theta_{W}$ was found to be -1.25 K. An important point to note here is that the poly-crystalline averaged $\theta_{W}$ remains negative at various temperature ranges used for fitting. This clearly indicates the dominance of AFM interactions in the system. The fitting parameters are listed in Sec. 2 of SI \cite{supply}. The effective magnetic moments along different directions are determined using the equation $\mu_{eff}$ = $\sqrt{8C}$. $\mu_{eff}$ from the polycrystalline average was found to be  $\sim$ 3.7 $\mu_B$, which is in excellent agreement with the theoretical value of the free Nd$^{3+}$ ion, which is approximately 3.62 $\mu_B$.

Figures \ref{fig2}(c) and \ref{fig2}(d) illustrate the magnetization isotherms of NdGaSi single crystal when $B \parallel $ [001] and $B \parallel $ [100], respectively. When the magnetic field is applied along [001], the magnetization grows with increasing field, with a sharp transition at around 0.3 T, and saturates with a maximum moment of 2.9 $\mu_B$ at 2 K, which is close to the expected free ion saturation moment. The critical field of 0.3 T acts as an inflection point. At higher temperatures, a sharp jump in magnetization is observed at the inflection point, beyond which the \textit{M}(\textit{H}) increases smoothly with increasing \textit{B}. The magnetization becomes progressively harder to saturate at higher temperatures, owning to domination thermal fluctuations, which makes the magnetic domains harder to align at higher magnetic fields. The magnetization displays a linear dependency at over 10 K, indicating the paramagnetic regime. On the other hand, along [100], magnetization at 2 K rises monotonically with increasing magnetic field until a metamagnetic transition occurs at around a critical field $B_c$ = 7.89 T. The magnetization continues to increase following the transition, with no indication of reaching saturation. At 9 T, the maximum moment observed is 1.85 $\mu_B$. Fig S2 \cite{supply} depicts the \textit{dM/dB} vs \textit{B} isotherms, revealing that the critical field diminishes with increasing temperature and vanishes about 10 K. This behavior is similar to the spin-flop metamagnetic transitions observed in many rare earth compounds \cite{arantes2018structure,muthuselvam2019gd2te3,ram2023multiple}.

It is also noteworthy that a similar kind of magnetization behavior in the \textit{M}(\textit{H}) isotherms has been observed before for ErNi$_5$ \cite{zhang1994crystalline}, whereas, in the case of NdGaSi, we observe the in-plane metamagnetic transition at relatively low $B_c$, unlike, the former, where the transition was seen at an extremely large $B_c$ = 18 T. Hence, we conclude that the ab-plane is the hard plane in NdGaSi as opposed to an easy \textit{c}-axis, with a maximum moment of 2.9 $\mu_B$ per Nd atom.  The nature of magnetization in the studied system is, thus, unique, and a detailed study of neutron diffraction should be employed further to deepen our understanding of its complex magnetic structure.

\subsection{Specific heat}
\begin{figure}[t]
\centering
\includegraphics[width=0.48\textwidth]{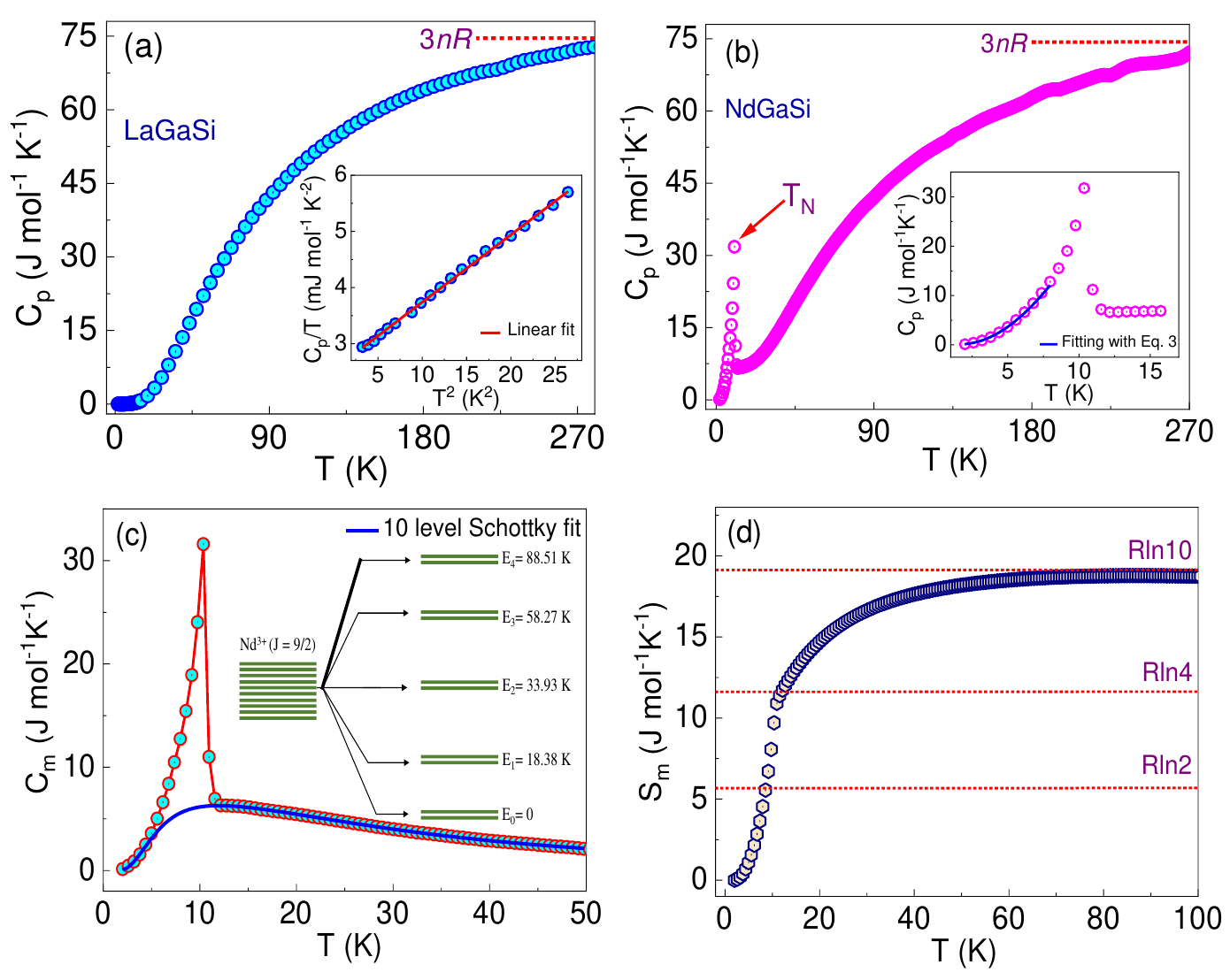}
\caption{Temperature-dependent molar specific heat of (a) single crystalline LaGaSi. The inset shows the linear fit of $C_p/T$ vs $T^2$.  (b) Molar specific heat of single crystalline NdGaSi. Inset shows the fitting of Eq. (3). (c) Magnetic component of specific heat as a function of temperature. The solid line represents the 10-level Schottky fit using the CEF scheme. Inset displays the energy level diagram according to the CEF scheme. (d) The magnetic entropy of NdGaSi as a function of temperature. }
\label{fig3}
\end{figure}

Figures \ref{fig3}(a) and \ref{fig3}(b) depict the specific heat of NdGaSi and nonmagnetic isostructural LaGaSi in a zero magnetic field. At high temperatures, the specific heat reaches its maximum value of approximately 72.84 J mol\textsuperscript{-1}K\textsuperscript{-1} for LaGaSi and 72.04 J mol\textsuperscript{-1}K\textsuperscript{-1} for NdGaSi, which are in good agreement with the predicted Dulong-Petit limit of 3\textit{nR} = 74.79 J mol\textsuperscript{-1}K\textsuperscript{-1}, where \textit{n} is the number of atoms per formula unit in the compound (\textit{n} = 3 for NdGaSi and LaGaSi), and \textit{R} = 8.31 J mol\textsuperscript{-1}K\textsuperscript{-1}. The specific heat of LaGaSi at a low-temperature regime is expressed by the equation

\begin{equation}
  C_p(T) = \gamma T + \beta T^3 
\end{equation}

where the first and second terms are the contributions due to the electronic and lattice contributions, respectively. The $\gamma$ and $\beta$ values are calculated by linear fitting the \textit{C\textsubscript{p}/T} vs \textit{T}\textsuperscript{2} data at a temperature region between 2 K and 5 K (Inset of Fig. \ref{fig3}(a)). We obtained  $\gamma$ = 2.539 mJ mol\textsuperscript{-1 }K\textsuperscript{-2} and Debye temperature ($\Theta_D$) = 365 K from the fitting parameters, which are in good agreement with the previously reported value of LaGaSi \cite{zhang2024magnetism}.

The specific heat data for NdGaSi (Fig. \ref{fig3}(b)) shows a sharp $\lambda$-like peak around 11 K, thus confirming the bulk nature of magnetic transition confirmed by our magnetic susceptibility data shown earlier. The specific heat of NdGaSi can be best expressed by the relation

\begin{equation}
  C_p(T) = \gamma T + \beta T^3 + \delta T^{3/2}e^{-\Delta/T}
\end{equation}

where the last term comes from the contribution due to the energy gap in the magnon spectrum \cite{gopal2012specific}. The value of $\beta$ was taken to be equal to that of LaGaSi assuming that the lattice contributions are similar. The fitting was performed between temperatures 2 K and 5 K (see inset of Fig. \ref{fig3}(b)) and we obtained the gap $\Delta$ = 6.8 K and $\gamma$ = 17.32 mJ mol$^{-1}$ K$^{-2}$, which is comparable to the order of some previously reported Nd$^{3+}$compounds \cite{campoy2006magnetoresistivity,szytula2007electronic}. 

Fig. \ref{fig3}(c) represents the magnetic contribution to the specific heat (C$_m$) which is obtained by subtracting the specific heat of the non-magnetic isostructural counterpart LaGaSi from that of NdGaSi. We observe a sharp peak at 11 K due to the magnetic transition and, additionally, a broad hump centered around 15 K. This hump is due to the well-known Schottky anomaly which is due to the splitting of the Nd$^{3+}$ atomic levels induced by crystalline electric field (CEF) and is quite common for rare earth compounds \cite{gopal2012specific}. The contribution of the Schottky anomaly to the heat capacity is expressed by the multilevel CEF equation

\begin{multline}
  C_{Sch}(T) = \\
   \left(\frac{R}{T^2}\right)\left[\frac{\sum_{i}^{ } g_{i}e^{\frac{-E_{i}}{T}} \sum_{i}^{ } g_{i}{E_{i}}^2e^{\frac{-E_{i}}{T}}-\left(\sum_{i}^{ } g_{i}E_ie^{\frac{-E_{i}}{T}} \right)^2}{\left(\sum_{i}^{ } g_{i}e^{\frac{-E_{i}}{T}} \right)^2}\right]
\end{multline}

where $E_i$ corresponds to $ith$ energy level (in K) having a degeneracy $g_i$ \cite{gopal2012specific}. Due to a tetragonal point group symmetry, the CEF potential will split the \textit{J} = 9/2 multiplets of the Krammers ion Nd$^{3+}$ into a ground state doublet and four other excited state doublets \cite{waltercef,allenspach1994magnetic}. To roughly understand the CEF scheme, we fitted the magnetic-specific heat data with Eqn. 4 and obtained a ground state doublet at $E_0$ = 0 K and four other excited state doublets at $E_1$ = 18.38 K, $E_2$ = 33.93 K, $E_3$ = 58.27 K, and $E_4$ = 88.51 K, in total five 2-fold degenerate levels, which justifies the energy splitting due to CEF in NdGaSi. This complete scheme is depicted in the inset of Fig. \ref{fig3}(c).

We have also calculated the magnetic entropy ($S_m$) of NdGaSi by using the following relation

\begin{equation}
  S_m(T) = \int\frac{C_m}{T}dT
\end{equation}

and is plotted as a function of temperature in Fig. \ref{fig3}(d). The $S_m$
reaches a value of around 9.3 J mol$^{-1}$K$^{-1}$ at $T_N$, lying between the ground state entropy $R$ln2 and $R$ln4. This indicates that the ground state doublet level is well below the magnetic transition and the first excited doublet level ($C_m$ = $R$ln4) is slightly higher than the transition, indicated by Fig. \ref{fig3}(d), thus requiring an additional entropy of 2.22 J mol$^{-1}$K$^{-1}$ to fully populate the first excited level located at around 11.8 K. This has been observed before in case of PrGaGe \cite{ram2023magnetic}. A total release of magnetic entropy at around 90 K is observed to be 18.78 J mol$^{-1}$K$^{-1}$, which is a good agreement with the total entropy released by an Nd$^{3+}$ ion (\textit{R}ln10).
\begin{figure}[t]
\centering
\includegraphics[width=0.47\textwidth]{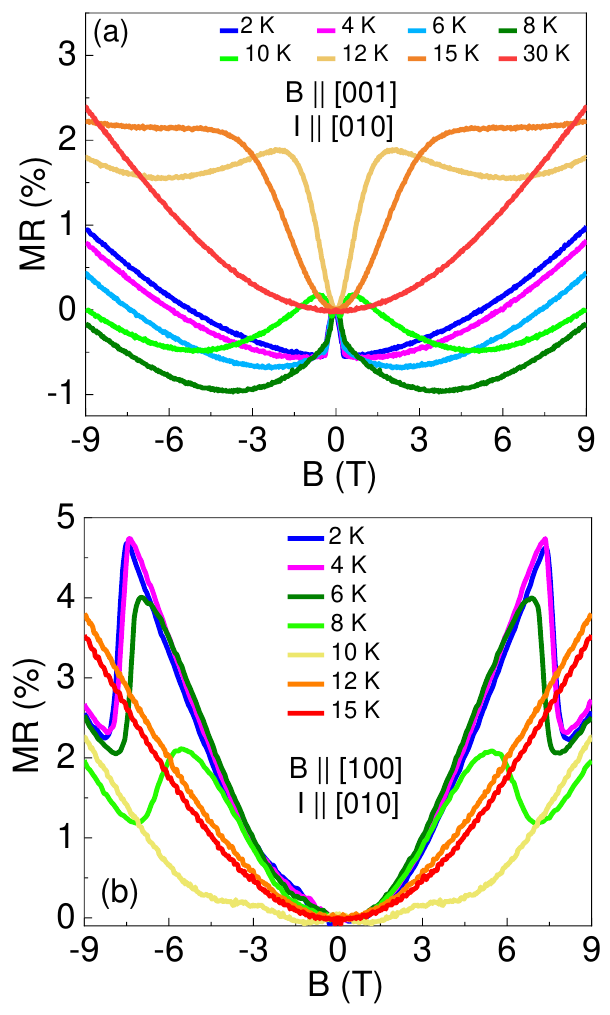}
\caption{Transverse magnetoresistance  (MR) measured as a function of the magnetic field with (a) $B \parallel $ [001] and (b) $B \parallel $ [100], with current $I \parallel$ [010] at different temperatures.}
\label{fig4}
\end{figure}

\begin{figure*}
\centering
\includegraphics[width=1.0\textwidth]{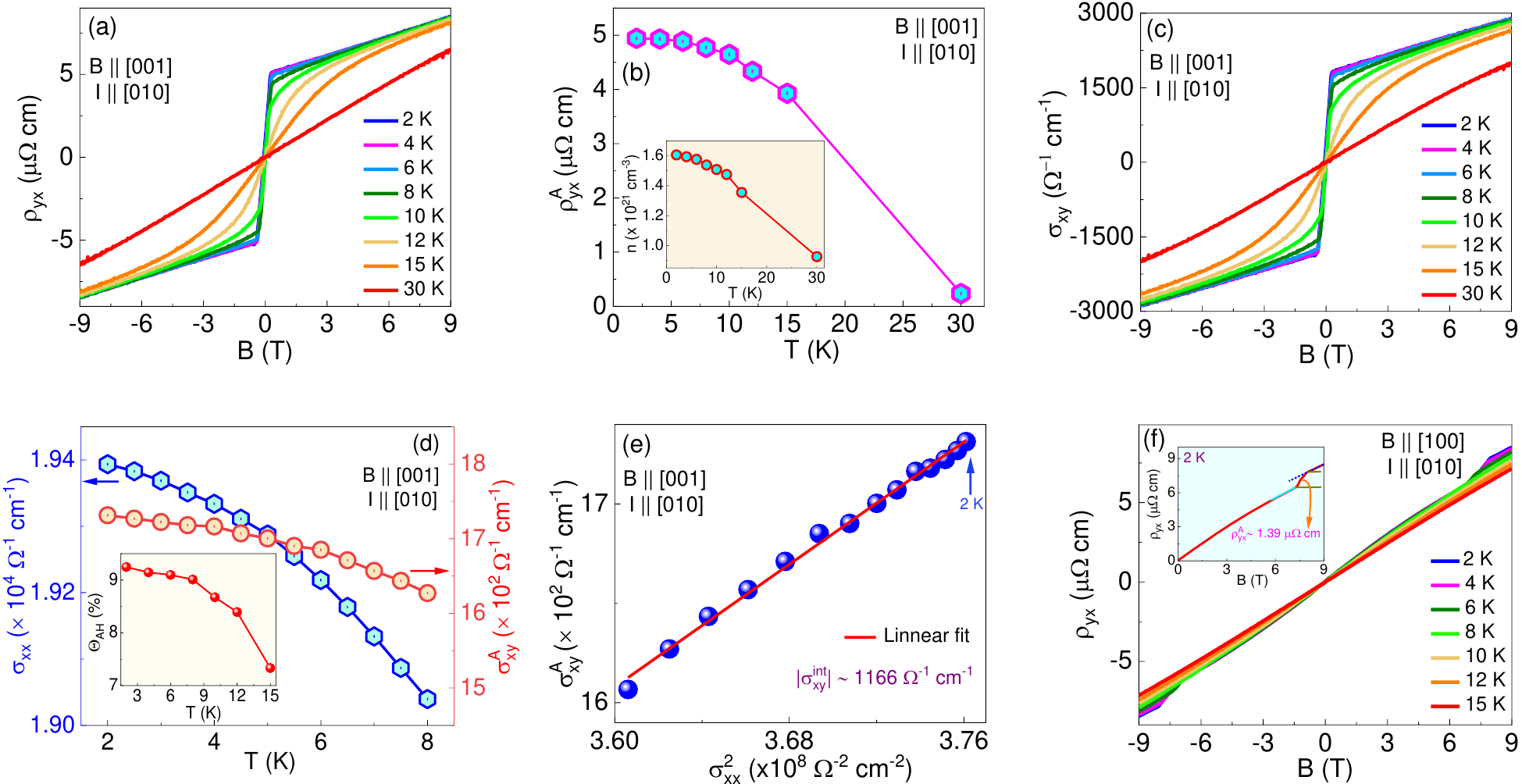}
\caption{(a) Magnetic field-dependent Hall resistivity ($\rho_{yx}$) at different temperatures ranging from 2 K to 30 K with $B \parallel$ [001] and $I \parallel$ [010]. (b) Temperature-dependent anomalous Hall resistivity $\rho^{A}_{yx}$. Inset: Temperature dependence of the carrier density ($n$) (c) Field-dependent Hall conductivity ($\sigma_{xy}$) at various temperatures. (d) Temperature dependence of longitudinal conductivity ($\sigma_{xx}$) and anomalous Hall conductivity ($\sigma^{A}_{xy}$). Inset displays the temperature dependence of anomalous Hall angle. (e) Linear fitting of anomalous Hall conductivity $\sigma^{A}_{xy}$ vs. $\sigma^{2}_{xx}$ curve. (f) The Hall resistivity with $B \parallel $ [100] and $I \parallel$ [010]. Inset represents the magnified view of the Hall resistivity at higher magnetic fields.} 
\label{fig5}
\vspace{-0.5cm}
\end{figure*}

\subsection{Magnetoresistance}
To gain more insight into the electronic transport properties, we measured the resistivity ($\rho_{xx}$) as a function of the magnetic field with the field applied along [001] and [100], as shown in Fig. \ref{fig4}(a) and \ref{fig4}(b), respectively. The current is applied along [010] for both scenarios. The magnetoresistance (MR) is calculated using the formula MR (\%) $= \frac{\rho_{xx}(B) - \rho_{xx}(0)}{\rho_{xx}(0)}\times100 $, where $\rho_{xx}(B)$ and $\rho_{xx}(0)$ correspond to the resistivities measured at magnetic field $B$ and zero fields, respectively. For the sample with $B \parallel $ [001] and $I \parallel$ [010], at 2 K, there is a sharp drop in MR at around 0.3 T, akin to the magnetic transition observed at the isothermal magnetization curve (see Fig. \ref{fig2}(c)). On increasing the field, the nature of MR changes, where we observe a positive increase with the field up to 9 T due to the dominance of orbital contribution due to Lorentz force. At higher temperatures, the MR drops at 0.3 T, the inflection point corresponding to the magnetization isotherm, but becomes progressively negative on increasing the magnetic field. The progression of negative MR even above 0.3 T is also consistent with the magnetization data where we see a region of slow increase of magnetization to a saturation value. At higher fields, a parabolic nature due to the dominance of orbital contribution is observed. Above the ordering regime beyond $T_{N}$, the nature of MR changes completely, as it starts to display a positive nature with increasing field, owning to the complete breakdown of the magnetic ordering and the establishment of the dominance of the orbital effect due to Lorentz force.

Figure \ref{fig4}(b) depicts the MR measured with $B \parallel $ [100] and $I\parallel $ [010]. At 2 K, we observe a steady increase in MR with increasing \textit{B,} followed by an abrupt decrease, implying the commencement of the metamagnetic transition, as observed for the \textit{M}(\textit{B}) curve above (see Fig. \ref{fig2}(d)). A similar phenomenon has been observed in other metamagnetic compounds \cite{hossain2000antiferromagnetic,ram2023multiple}. With the rising temperature, the critical field of MR decreases, following the magnetization curve. Beyond 11 K, we observe that the orbital term begins to dominate the MR, gradually modifying its nature to be a complete parabolic one.

\subsection{Anomalous Hall effect}

To explore the anisotropic anomalous transport properties of NdAlSi, we have acquired the Hall resistivity ($\rho_{yx}$) data in both directions. Figure \ref{fig5}(a) shows the field-dependent $\rho_{yx}$ curve measured with $I \parallel $ [010] and $B \parallel $ [001] at various temperatures. At 30 K, somewhat higher than $T_N$, the $\rho_{yx}$ is linear, as expected for an ordinary conductor. When the temperature is lowered below $T_N$, an additional contribution appears in the Hall resistivity. This anomalous behavior is observed up to $\sim$ 0.3 T, beyond which $\rho_{yx}$ exhibits a subtle linear field dependence up to 9 T. The similarity between the shape of the \textit{M}(\textit{B}) and $\rho_{yx} (B)$ curves below $T_N$ clearly indicates the presence of AHE in the present compound. Conventionally, the total Hall resistivity in a ferromagnet can be represented as $\rho_{yx} (B)$ = $\rho_{yx} ^O$ + $\rho_{yx} ^A$ = $R_0B$ + $R_s$$\mu_0$$M$, where $\rho_{yx} ^O$ and $\rho_{yx} ^A$ are the ordinary and anomalous contributions to the total Hall resistivity, with $R_0$ and $R_s$ being the ordinary and anomalous Hall coefficients, respectively. The linear fit of the $\rho_{yx}$ vs. \textit{B} curve in the high-field region yields the values of $\rho_{yx} ^A$ and $R_0$. The slope and the \textit{y}-axis intercept of the linear fit correspond to $R_0$ and $\rho_{yx}^A$, respectively. In the single-band model, the carrier density $n$ can be approximated using $R_0$ ($n$ = 1/$R_0e$, where \textit{e} represents the charge of an electron). Figure \ref{fig5}(b) displays the temperature dependence of $\rho_{yx}^A$ and $n$, which is almost constant below $T_N$ and falls off rapidly above it. At 2 K, a carrier density of $\sim$ 1.6 $\times$ 10$^{21}$ is obtained; the positive values of $n$ suggest that holes are the majority of charge carriers throughout the temperature range. 

In order to understand the microscopic mechanism responsible for the observed AHE in NdAlSi, we need to look into the variation of AHC $\sigma_{xy}^A$ with longitudinal conductivity $\sigma_{xx}$. Consequently, we first determine the Hall conductivity $\sigma_{xy}$ using the tensor conversion formula:

\begin{equation}
\sigma_{xy} = \frac{\rho_{yx}}{\rho_{xx}^2 + \rho_{yx}^2}
\end{equation}

Figure \ref{fig5}(c) represents the field-dependent $\sigma_{xy}$ with $B \parallel $ [001] and $I \parallel$ [010] at different temperatures. Using zero-field extrapolation of high-field $\sigma_{xy}$ data on the \textit{y}-axis, we have extracted the AHC $\sigma_{xy}^A$. Remarkably, we have observed a giant AHC of $\sim$ 1730 $\Omega^{-1}$ cm$^{-1}$ at 2 K (see Fig. \ref{fig5}(d)), which is much larger than most of the previously reported AFMs \cite{shekhar2018anomalous,kotegawa2023large}. Likewise, we have noticed an exceptionally large anomalous Hall angle (AHA) of $\sim$ 9.3 \% at 2 K as shown in the inset of Fig. \ref{fig5}(d), which is exceedingly uncommon in both FMs and AFMs \cite{kim2018large,chatterjee2023nodal}. To confirm the consistency of the giant AHC value obtained, we have measured the Hall resistivity $\sigma_{xy}^A$ at 2 K with $B \parallel $ [001] and $I \parallel $ [010] for two additional NdGaSi single crystals, as described in Sec. S4 of ref. \cite{supply}. $\sigma_{xy}^A$ arises mainly from three primary mechanisms: the intrinsic Karplus-Luttinger (KL) mechanism, extrinsic skew scattering, and side jump mechanisms \cite{nagaosa2010anomalous}. The intrinsic KL contribution to $\sigma_{xy}^A$ is expected to be temperature-independent, as it is entirely dependent on the Berry curvature of the electronic bands and is not affected by the scattering mechanisms \cite{karplus1954hall,jungwirth2002anomalous}. Figure \ref{fig5}(d) illustrates that below $T_N$, the $\sigma_{xy}^A$ has a very weak temperature dependence, while the $\sigma_{xx}$ is largely temperature-dependent, suggesting that the AHE in NdAlSi predominantly originates from the intrinsic Berry phase mechanism. It is also important to note that the longitudinal conductivity $\sigma_{xx}$ of NdAlSi falls between the bad metal and the clean limit (i.e., 10$^4$ $\leq$ $\sigma_{xx}$ $\leq$ 10$^6$). The RRR value for NdAlSi is $\leq$ 2 and the AHE occurs at very low temperatures ($<$ 11 K), which renders the role of phonon scattering nearly ineffectual and is dominated by the residual resistivity at low-temperature $\rho_{xx0}$. In such a scenario, the AHC can be scaled by the following empirical formula \cite{tian2009proper,hou2015multivariable,yang2020transition}

 \begin{equation}
   \sigma_{xy}^A = (\alpha\sigma_{xx0}^{-1} + \beta\sigma_{xx0}^{-2})\sigma_{xx}^2 + \gamma
\end{equation}

where $\sigma_{xy}^A $ is the AHC, $\sigma_{xx0}$ is the residual longitudinal conductivity and the coefficients $\alpha$, $\beta$, and $\gamma$ correspond to the skew-scattering, side jump, and the intrinsic Berry phase contributions to the $\sigma_{yx}^A$, respectively. In Fig. \ref{fig5}(e), the $\gamma$ (or $\sigma_{xy}^{int}$) is determined by employing a linear fitting between $\sigma_{xy}^A$ and $\sigma_{xx}^2$. It has been observed that the magnitude of the intrinsic Berry phase contribution ($\sigma_{xy}^{int}$) is $\sim$ 1166 $\Omega^{-1}$ cm$^{-1}$, with a total AHC of approximately 1730 $\Omega^{-1}$ cm$^{-1}$ at 2 K, further suggesting that the AHE in NdAlSi is dominated by the intrinsic Berry phase mechanism. Here, it is also important to highlight that we require a sufficient number of data points in the temperature range from 2 to 8 K (below the $T_N$) to perform a proper fit between $\sigma_{xy}^A$ and $\sigma_{xx}^2$ in Fig. \ref{fig5}(e). For this reason, we have plotted the Hall resistivity vs. temperature curve at five different magnetic fields well above saturation and obtained the $\rho_{yx}^A$ data points at a separation of 0.5 K by linear fitting the $\rho_{yx}$ vs. $B$ data at fixed temperatures (see the details in Sec. 3 of SI \cite{supply}). This allowed us to gather a large number of data points from 2 to 8 K. A similar way has also been employed to determine a denser number of $\rho_{yx}^A$ data points from 2 to 8 K in Fig. \ref{fig5}(d).

Figure \ref{fig5}(f) displays the field-dependent Hall resistivity $\rho_{yx}$ at various temperatures with $B \parallel $ [100] and $I \parallel $ [010] Amazingly, we see an anomalous Hall response with $B \parallel $ [100] below $T_N$, in contrast to the other member of the family CeGaSi, where AHE is only present for one direction \cite{gong2024anomalous}. However, the observed anomalous Hall resistivity $\rho_{yx}^A$ with $B \parallel $ [100] is about three times smaller than $\rho_{yx}^A$ with $B \parallel $ [001] as shown in the inset of Fig. \ref{fig5}(f), indicating a complex nature of magnetism in both directions. The $\rho_{yx}^A$ is derived in this case by taking the height of the metamagnetic transition in the $\rho_{yx}$ vs. \textit{B} data (see inset of Fig. \ref{fig5}(f)), as adopted for other related compounds \cite{ram2023multiple,zhou2023metamagnetic}. We have also estimated the AHC using the formula $\sigma_{xy}^A = \rho_{yx}^A/\rho_{xx}^2$ and found a large AHC of around 490 $\Omega^{-1}$ cm$^{-1}$ at 2 K, which is comparable to the large AHC value observed in Mn$_3$Sn and Mn$_3$Ge \cite{nakatsuji2015large,nayak2016large}.

\begin{figure}[t]
\centering
\includegraphics[width=0.45\textwidth]{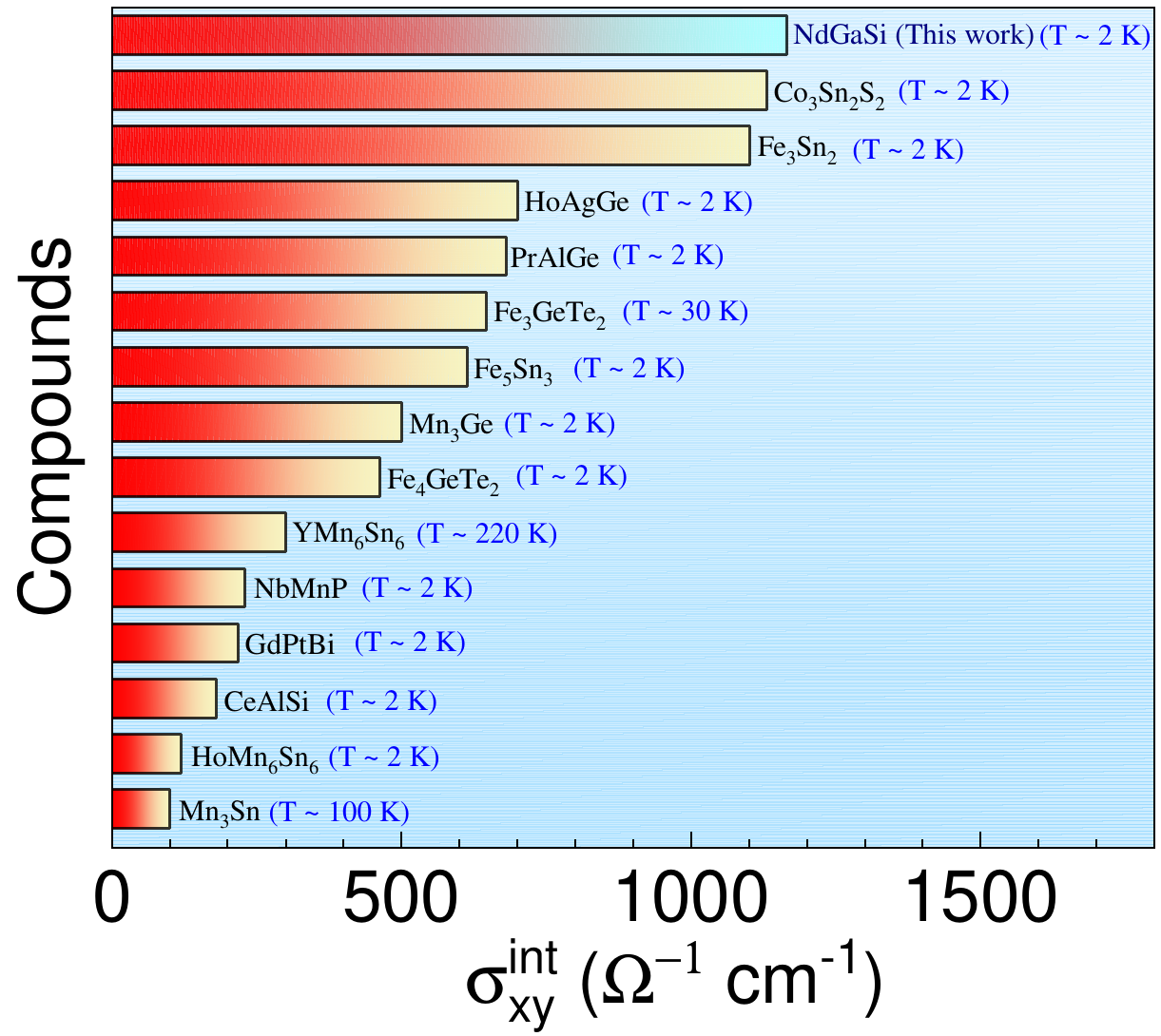}
\caption{Comparison of intrinsic anomalous Hall conductivity ($\sigma^{int}_{xy}$) value of the studied system (NdGaSi) with previously reported ferromagnetic (FM) and antiferromagnetic (AFM) compounds.}
\label{fig6}
\end{figure}

Finally, to summarize our findings in NdAlSi, we have compared our intrinsic AHC ($\sigma_{xy}^{int}$) value to previously reported componds in Fig. \ref{fig6}. NdAlSi shows an intrinsic AHC value of $\sim$ 1166 $\Omega^{-1}$ cm$^{-1}$ at 2 K, with $B \parallel $ [001] and $I \parallel $ [010], exceeding all other known AFM systems and is comparable to FM systems \cite{liu2018giant,ye2018massive,yang2020transition,wang2017anisotropic,li2020large,bera2023anomalous,alam2023sign,roychowdhury2024enhancement,nayak2016large,wang2021field,arai2024intrinsic,suzuki2016large,nakatsuji2015large,zeng2022large}.

\section{CONCLUSION}
\vspace{3mm}
In conclusion, we have thoroughly studied the magnetic, thermodynamic, and transport properties of NdGaSi single crystal. The temperature-dependent magnetic susceptibility $\chi (T)$ and longitudinal resistivity $\rho_{xx}(T)$ suggest the N\'{e}el temperature is at $T_N \sim$ 11 K. The bulk character of the magnetic ordering is further supported by the specific heat data, which exhibit a large peak near the $T_N$. The magnetic measurements also indicate the presence of magnetic anisotropy, which exhibits ferromagnetic ordering along [001] and antiferromagnetic ordering along [100]. The MR suggests a spin flop-induced metamagnetic behavior along the hard axis and a spin quenching-induced negative trend along the easy axis. Moreover, we have observed a giant AHC of approximately 1730 $\Omega^{-1}$ cm$^{-1}$ with an intrinsic contribution of $\sim$ 1165 $\Omega^{-1}$ cm$^{-1}$ and a large AHA of $\sim$ 9.3\% at 2 K, with $B \parallel $ [001], which is significantly larger than benchmark AFM materials like Mn$_3$Sn and Mn$_3$Ge. Additionally, at 2 K, we have also found an AHC of about 490 $\Omega^{-1}$ cm$^{-1}$ with $B \parallel $ [100]. Our MR and Hall resistivity data in two directions demonstrate anisotropic magnetotransport characteristics in NdGaSi. Additional theoretical calculations and neutron diffraction studies are necessary to fully understand the electronic band structure and magnetic characteristics of NdGaSi.

\section*{ACKNOWLEDGEMENTS}
\vspace{3mm}
NK acknowledges DST for financial support through Grant Sanction No. CRG/2021/002747 and Max Planck Society for funding under the Max Planck-India partner group project. This research project made use of the instrumentation facility provided by the Technical Research Centre (TRC) at the S.N. Bose National Centre for Basic Sciences, under the Department of Science and Technology, Government of India.


\begin{thebibliography}{58}%
\makeatletter
\providecommand \@ifxundefined [1]{%
 \@ifx{#1\undefined}
}%
\providecommand \@ifnum [1]{%
 \ifnum #1\expandafter \@firstoftwo
 \else \expandafter \@secondoftwo
 \fi
}%
\providecommand \@ifx [1]{%
 \ifx #1\expandafter \@firstoftwo
 \else \expandafter \@secondoftwo
 \fi
}%
\providecommand \natexlab [1]{#1}%
\providecommand \enquote  [1]{``#1''}%
\providecommand \bibnamefont  [1]{#1}%
\providecommand \bibfnamefont [1]{#1}%
\providecommand \citenamefont [1]{#1}%
\providecommand \@href[1]{\@@startlink{#1}\@@href}%
\providecommand \@@href[1]{\endgroup#1\@@endlink}%
\providecommand \@sanitize@url [0]{\catcode `\\12\catcode `\$12\catcode `\&12\catcode `\#12\catcode `\^12\catcode `\_12\catcode `\%12\relax}%
\providecommand \@@startlink[1]{}%
\providecommand \@@endlink[0]{}%
\providecommand \@url [1]{\endgroup\@href {#1}{\urlprefix }}%
\providecommand \urlprefix  [0]{URL }%
\providecommand \doibase [0]{https://doi.org/}%
\providecommand \selectlanguage [0]{\@gobble}%
\providecommand \bibinfo  [0]{\@secondoftwo}%
\providecommand \bibfield  [0]{\@secondoftwo}%
\providecommand \translation [1]{[#1]}%
\providecommand \BibitemOpen [0]{}%
\providecommand \bibitemStop [0]{}%
\providecommand \bibitemNoStop [0]{.\EOS\space}%
\providecommand \EOS [0]{\spacefactor3000\relax}%
\providecommand \BibitemShut  [1]{\csname bibitem#1\endcsname}%
\let\auto@bib@innerbib\@empty
\bibitem [{\citenamefont {Steglich}\ \emph {et~al.}(1979)\citenamefont {Steglich}, \citenamefont {Aarts}, \citenamefont {Bredl}, \citenamefont {Lieke}, \citenamefont {Meschede}, \citenamefont {Franz},\ and\ \citenamefont {Sch{\"a}fer}}]{steglich1979superconductivity}%
  \BibitemOpen
  \bibfield  {author} {\bibinfo {author} {\bibfnamefont {F.}~\bibnamefont {Steglich}}, \bibinfo {author} {\bibfnamefont {J.}~\bibnamefont {Aarts}}, \bibinfo {author} {\bibfnamefont {C.}~\bibnamefont {Bredl}}, \bibinfo {author} {\bibfnamefont {W.}~\bibnamefont {Lieke}}, \bibinfo {author} {\bibfnamefont {D.}~\bibnamefont {Meschede}}, \bibinfo {author} {\bibfnamefont {W.}~\bibnamefont {Franz}},\ and\ \bibinfo {author} {\bibfnamefont {H.}~\bibnamefont {Sch{\"a}fer}},\ }\bibfield  {title} {\bibinfo {title} {Superconductivity in the presence of strong Pauli paramagnetism: \ch{CeCu2Si2}},\ } {\bibfield  {journal} {\bibinfo  {journal} {Physical Review Letters}\ }\textbf {\bibinfo {volume} {43}},\ \bibinfo {pages} {1892} (\bibinfo {year} {1979})}\BibitemShut {NoStop}%
\bibitem [{\citenamefont {Mathur}\ \emph {et~al.}(1998)\citenamefont {Mathur}, \citenamefont {Grosche}, \citenamefont {Julian}, \citenamefont {Walker}, \citenamefont {Freye}, \citenamefont {Haselwimmer},\ and\ \citenamefont {Lonzarich}}]{mathur1998magnetically}%
  \BibitemOpen
  \bibfield  {author} {\bibinfo {author} {\bibfnamefont {N.}~\bibnamefont {Mathur}}, \bibinfo {author} {\bibfnamefont {F.}~\bibnamefont {Grosche}}, \bibinfo {author} {\bibfnamefont {S.}~\bibnamefont {Julian}}, \bibinfo {author} {\bibfnamefont {I.}~\bibnamefont {Walker}}, \bibinfo {author} {\bibfnamefont {D.}~\bibnamefont {Freye}}, \bibinfo {author} {\bibfnamefont {R.}~\bibnamefont {Haselwimmer}},\ and\ \bibinfo {author} {\bibfnamefont {G.}~\bibnamefont {Lonzarich}},\ }\bibfield  {title} {\bibinfo {title} {Magnetically mediated superconductivity in heavy fermion compounds},\ } {\bibfield  {journal} {\bibinfo  {journal} {Nature}\ }\textbf {\bibinfo {volume} {394}},\ \bibinfo {pages} {39} (\bibinfo {year} {1998})}\BibitemShut {NoStop}%
\bibitem [{\citenamefont {Sacchetti}\ \emph {et~al.}(2007)\citenamefont {Sacchetti}, \citenamefont {Arcangeletti}, \citenamefont {Perucchi}, \citenamefont {Baldassarre}, \citenamefont {Postorino}, \citenamefont {Lupi}, \citenamefont {Ru}, \citenamefont {Fisher},\ and\ \citenamefont {Degiorgi}}]{sacchetti2007pressure}%
  \BibitemOpen
  \bibfield  {author} {\bibinfo {author} {\bibfnamefont {A.}~\bibnamefont {Sacchetti}}, \bibinfo {author} {\bibfnamefont {E.}~\bibnamefont {Arcangeletti}}, \bibinfo {author} {\bibfnamefont {A.}~\bibnamefont {Perucchi}}, \bibinfo {author} {\bibfnamefont {L.}~\bibnamefont {Baldassarre}}, \bibinfo {author} {\bibfnamefont {P.}~\bibnamefont {Postorino}}, \bibinfo {author} {\bibfnamefont {S.}~\bibnamefont {Lupi}}, \bibinfo {author} {\bibfnamefont {N.}~\bibnamefont {Ru}}, \bibinfo {author} {\bibfnamefont {I.~R.}\ \bibnamefont {Fisher}},\ and\ \bibinfo {author} {\bibfnamefont {L.}~\bibnamefont {Degiorgi}},\ }\bibfield  {title} {\bibinfo {title} {Pressure dependence of the charge-density-wave gap in rare-earth tritellurides},\ }{\bibfield  {journal} {\bibinfo  {journal} {Physical Review Letters}\ }\textbf {\bibinfo {volume} {98}},\ \bibinfo {pages} {026401} (\bibinfo {year} {2007})}\BibitemShut {NoStop}%
\bibitem [{\citenamefont {Brouet}\ \emph {et~al.}(2008)\citenamefont {Brouet}, \citenamefont {Yang}, \citenamefont {Zhou}, \citenamefont {Hussain}, \citenamefont {Moore}, \citenamefont {He}, \citenamefont {Lu}, \citenamefont {Shen}, \citenamefont {Laverock}, \citenamefont {Dugdale} \emph {et~al.}}]{brouet2008angle}%
  \BibitemOpen
  \bibfield  {author} {\bibinfo {author} {\bibfnamefont {V.}~\bibnamefont {Brouet}}, \bibinfo {author} {\bibfnamefont {W.}~\bibnamefont {Yang}}, \bibinfo {author} {\bibfnamefont {X.}~\bibnamefont {Zhou}}, \bibinfo {author} {\bibfnamefont {Z.}~\bibnamefont {Hussain}}, \bibinfo {author} {\bibfnamefont {R.}~\bibnamefont {Moore}}, \bibinfo {author} {\bibfnamefont {R.}~\bibnamefont {He}}, \bibinfo {author} {\bibfnamefont {D.}~\bibnamefont {Lu}}, \bibinfo {author} {\bibfnamefont {Z.}~\bibnamefont {Shen}}, \bibinfo {author} {\bibfnamefont {J.}~\bibnamefont {Laverock}}, \bibinfo {author} {\bibfnamefont {S.}~\bibnamefont {Dugdale}}, \emph {et~al.},\ }\bibfield  {title} {\bibinfo {title} {Angle-resolved photoemission study of the evolution of band structure and charge density wave properties in \ch{RTe3} ({R} = {Y}, {L}a, {C}e, {S}m, {G}d, {T}b, and {D}y)},\ }{\bibfield  {journal} {\bibinfo  {journal} {Physical Review B}\ }\textbf {\bibinfo {volume} {77}},\
  \bibinfo {pages} {235104} (\bibinfo {year} {2008})}\BibitemShut {NoStop}%
\bibitem [{\citenamefont {Gaudet}\ \emph {et~al.}(2021)\citenamefont {Gaudet}, \citenamefont {Yang}, \citenamefont {Baidya}, \citenamefont {Lu}, \citenamefont {Xu}, \citenamefont {Zhao}, \citenamefont {Rodriguez-Rivera}, \citenamefont {Hoffmann}, \citenamefont {Graf}, \citenamefont {Torchinsky} \emph {et~al.}}]{gaudet2021weyl}%
  \BibitemOpen
  \bibfield  {author} {\bibinfo {author} {\bibfnamefont {J.}~\bibnamefont {Gaudet}}, \bibinfo {author} {\bibfnamefont {H.~Y.}\ \bibnamefont {Yang}}, \bibinfo {author} {\bibfnamefont {S.}~\bibnamefont {Baidya}}, \bibinfo {author} {\bibfnamefont {B.}~\bibnamefont {Lu}}, \bibinfo {author} {\bibfnamefont {G.}~\bibnamefont {Xu}}, \bibinfo {author} {\bibfnamefont {Y.}~\bibnamefont {Zhao}}, \bibinfo {author} {\bibfnamefont {J.~A.}\ \bibnamefont {Rodriguez-Rivera}}, \bibinfo {author} {\bibfnamefont {C.~M.}\ \bibnamefont {Hoffmann}}, \bibinfo {author} {\bibfnamefont {D.~E.}\ \bibnamefont {Graf}}, \bibinfo {author} {\bibfnamefont {D.~H.}\ \bibnamefont {Torchinsky}}, \emph {et~al.},\ }\bibfield  {title} {\bibinfo {title} {Weyl-mediated helical magnetism in \ch{NdAlSi}},\ }{\bibfield  {journal} {\bibinfo  {journal} {Nature Materials}\ }\textbf {\bibinfo {volume} {20}},\ \bibinfo {pages} {1650} (\bibinfo {year} {2021})}\BibitemShut {NoStop}%
\bibitem [{\citenamefont {Cheng}\ \emph {et~al.}(2024)\citenamefont {Cheng}, \citenamefont {Yan}, \citenamefont {Shi}, \citenamefont {Lou}, \citenamefont {Fedorov}, \citenamefont {Behnami}, \citenamefont {Yuan}, \citenamefont {Yang}, \citenamefont {Wang}, \citenamefont {Cheng} \emph {et~al.}}]{cheng2024tunable}%
  \BibitemOpen
  \bibfield  {author} {\bibinfo {author} {\bibfnamefont {E.}~\bibnamefont {Cheng}}, \bibinfo {author} {\bibfnamefont {L.}~\bibnamefont {Yan}}, \bibinfo {author} {\bibfnamefont {X.}~\bibnamefont {Shi}}, \bibinfo {author} {\bibfnamefont {R.}~\bibnamefont {Lou}}, \bibinfo {author} {\bibfnamefont {A.}~\bibnamefont {Fedorov}}, \bibinfo {author} {\bibfnamefont {M.}~\bibnamefont {Behnami}}, \bibinfo {author} {\bibfnamefont {J.}~\bibnamefont {Yuan}}, \bibinfo {author} {\bibfnamefont {P.}~\bibnamefont {Yang}}, \bibinfo {author} {\bibfnamefont {B.}~\bibnamefont {Wang}}, \bibinfo {author} {\bibfnamefont {~J.-G.}\ \bibnamefont {Cheng}}, \emph {et~al.},\ }\bibfield  {title} {\bibinfo {title} {Tunable positions of Weyl nodes via magnetism and pressure in the ferromagnetic Weyl semimetal \ch{CeAlSi}},\ }{\bibfield  {journal} {\bibinfo  {journal} {Nature Communications}\ }\textbf {\bibinfo {volume} {15}},\ \bibinfo {pages} {1467} (\bibinfo {year} {2024})}\BibitemShut {NoStop}%
\bibitem [{\citenamefont {Chang}\ \emph {et~al.}(2018)\citenamefont {Chang}, \citenamefont {Singh}, \citenamefont {Xu}, \citenamefont {Bian}, \citenamefont {Huang}, \citenamefont {Hsu}, \citenamefont {Belopolski}, \citenamefont {Alidoust}, \citenamefont {Sanchez}, \citenamefont {Zheng} \emph {et~al.}}]{chang2018magnetic}%
  \BibitemOpen
  \bibfield  {author} {\bibinfo {author} {\bibfnamefont {G.}~\bibnamefont {Chang}}, \bibinfo {author} {\bibfnamefont {B.}~\bibnamefont {Singh}}, \bibinfo {author} {\bibfnamefont {S.-Y.}\ \bibnamefont {Xu}}, \bibinfo {author} {\bibfnamefont {G.}~\bibnamefont {Bian}}, \bibinfo {author} {\bibfnamefont {S.-M.}\ \bibnamefont {Huang}}, \bibinfo {author} {\bibfnamefont {C.-H.}\ \bibnamefont {Hsu}}, \bibinfo {author} {\bibfnamefont {I.}~\bibnamefont {Belopolski}}, \bibinfo {author} {\bibfnamefont {N.}~\bibnamefont {Alidoust}}, \bibinfo {author} {\bibfnamefont {D.~S.}\ \bibnamefont {Sanchez}}, \bibinfo {author} {\bibfnamefont {H.}~\bibnamefont {Zheng}}, \emph {et~al.},\ }\bibfield  {title} {\bibinfo {title} {Magnetic and noncentrosymmetric {W}eyl fermion semimetals in the \ch{RAlGe} family of compounds ({R}= rare earth)},\ }{\bibfield  {journal} {\bibinfo  {journal} {Physical Review B}\ }\textbf {\bibinfo {volume} {97}},\ \bibinfo {pages} {041104} (\bibinfo {year} {2018})}\BibitemShut {NoStop}%
\bibitem [{\citenamefont {Suzuki}\ \emph {et~al.}(2019)\citenamefont {Suzuki}, \citenamefont {Savary}, \citenamefont {Liu}, \citenamefont {Lynn}, \citenamefont {Balents},\ and\ \citenamefont {Checkelsky}}]{suzuki2019singular}%
  \BibitemOpen
  \bibfield  {author} {\bibinfo {author} {\bibfnamefont {T.}~\bibnamefont {Suzuki}}, \bibinfo {author} {\bibfnamefont {L.}~\bibnamefont {Savary}}, \bibinfo {author} {\bibfnamefont {J.-P.}\ \bibnamefont {Liu}}, \bibinfo {author} {\bibfnamefont {J.~W.}\ \bibnamefont {Lynn}}, \bibinfo {author} {\bibfnamefont {L.}~\bibnamefont {Balents}},\ and\ \bibinfo {author} {\bibfnamefont {J.~G.}\ \bibnamefont {Checkelsky}},\ }\bibfield  {title} {\bibinfo {title} {Singular angular magnetoresistance in a magnetic nodal semimetal},\ } {\bibfield  {journal} {\bibinfo  {journal} {Science}\ }\textbf {\bibinfo {volume} {365}},\ \bibinfo {pages} {377} (\bibinfo {year} {2019})}\BibitemShut {NoStop}%
\bibitem [{\citenamefont {Yang}\ \emph {et~al.}(2021)\citenamefont {Yang}, \citenamefont {Singh}, \citenamefont {Gaudet}, \citenamefont {Lu}, \citenamefont {Huang}, \citenamefont {Chiu}, \citenamefont {Huang}, \citenamefont {Wang}, \citenamefont {Bahrami}, \citenamefont {Xu} \emph {et~al.}}]{yang2021noncollinear}%
  \BibitemOpen
  \bibfield  {author} {\bibinfo {author} {\bibfnamefont {H.-Y.}\ \bibnamefont {Yang}}, \bibinfo {author} {\bibfnamefont {B.}~\bibnamefont {Singh}}, \bibinfo {author} {\bibfnamefont {J.}~\bibnamefont {Gaudet}}, \bibinfo {author} {\bibfnamefont {B.}~\bibnamefont {Lu}}, \bibinfo {author} {\bibfnamefont {C.-Y.}\ \bibnamefont {Huang}}, \bibinfo {author} {\bibfnamefont {W.-C.}\ \bibnamefont {Chiu}}, \bibinfo {author} {\bibfnamefont {S.-M.}\ \bibnamefont {Huang}}, \bibinfo {author} {\bibfnamefont {B.}~\bibnamefont {Wang}}, \bibinfo {author} {\bibfnamefont {F.}~\bibnamefont {Bahrami}}, \bibinfo {author} {\bibfnamefont {B.}~\bibnamefont {Xu}}, \emph {et~al.},\ }\bibfield  {title} {\bibinfo {title} {Noncollinear ferromagnetic {W}eyl semimetal with anisotropic anomalous {H}all effect},\ } {\bibfield  {journal} {\bibinfo  {journal} {Physical Review B}\ }\textbf {\bibinfo {volume} {103}},\ \bibinfo {pages} {115143} (\bibinfo {year} {2021})}\BibitemShut {NoStop}%
\bibitem [{\citenamefont {Puphal}\ \emph {et~al.}(2020)\citenamefont {Puphal}, \citenamefont {Pomjakushin}, \citenamefont {Kanazawa}, \citenamefont {Ukleev}, \citenamefont {Gawryluk}, \citenamefont {Ma}, \citenamefont {Naamneh}, \citenamefont {Plumb}, \citenamefont {Keller}, \citenamefont {Cubitt} \emph {et~al.}}]{puphal2020topological}%
  \BibitemOpen
  \bibfield  {author} {\bibinfo {author} {\bibfnamefont {P.}~\bibnamefont {Puphal}}, \bibinfo {author} {\bibfnamefont {V.}~\bibnamefont {Pomjakushin}}, \bibinfo {author} {\bibfnamefont {N.}~\bibnamefont {Kanazawa}}, \bibinfo {author} {\bibfnamefont {V.}~\bibnamefont {Ukleev}}, \bibinfo {author} {\bibfnamefont {D.~J.}\ \bibnamefont {Gawryluk}}, \bibinfo {author} {\bibfnamefont {J.}~\bibnamefont {Ma}}, \bibinfo {author} {\bibfnamefont {M.}~\bibnamefont {Naamneh}}, \bibinfo {author} {\bibfnamefont {N.~C.}\ \bibnamefont {Plumb}}, \bibinfo {author} {\bibfnamefont {L.}~\bibnamefont {Keller}}, \bibinfo {author} {\bibfnamefont {R.}~\bibnamefont {Cubitt}}, \emph {et~al.},\ }\bibfield  {title} {\bibinfo {title} {Topological magnetic phase in the candidate {W}eyl semimetal \ch{CeAlGe}},\ }{\bibfield  {journal} {\bibinfo  {journal} {Physical Review Letters}\ }\textbf {\bibinfo {volume} {124}},\ \bibinfo {pages} {017202} (\bibinfo {year} {2020})}\BibitemShut {NoStop}%
\bibitem [{\citenamefont {Lyu}\ \emph {et~al.}(2020)\citenamefont {Lyu}, \citenamefont {Xiang}, \citenamefont {Mi}, \citenamefont {Zhao}, \citenamefont {Wang}, \citenamefont {Liu}, \citenamefont {Chen}, \citenamefont {Ren}, \citenamefont {Li},\ and\ \citenamefont {Sun}}]{lyu2020nonsaturating}%
  \BibitemOpen
  \bibfield  {author} {\bibinfo {author} {\bibfnamefont {M.}~\bibnamefont {Lyu}}, \bibinfo {author} {\bibfnamefont {J.}~\bibnamefont {Xiang}}, \bibinfo {author} {\bibfnamefont {Z.}~\bibnamefont {Mi}}, \bibinfo {author} {\bibfnamefont {H.}~\bibnamefont {Zhao}}, \bibinfo {author} {\bibfnamefont {Z.}~\bibnamefont {Wang}}, \bibinfo {author} {\bibfnamefont {E.}~\bibnamefont {Liu}}, \bibinfo {author} {\bibfnamefont {G.}~\bibnamefont {Chen}}, \bibinfo {author} {\bibfnamefont {Z.}~\bibnamefont {Ren}}, \bibinfo {author} {\bibfnamefont {G.}~\bibnamefont {Li}},\ and\ \bibinfo {author} {\bibfnamefont {P.}~\bibnamefont {Sun}},\ }\bibfield  {title} {\bibinfo {title} {Nonsaturating magnetoresistance, anomalous {H}all effect, and magnetic quantum oscillations in the ferromagnetic semimetal \ch{PrAlSi}},\ }{\bibfield  {journal} {\bibinfo  {journal} {Physical Review B}\ }\textbf {\bibinfo {volume} {102}},\ \bibinfo {pages} {085143} (\bibinfo {year} {2020})}\BibitemShut {NoStop}%
\bibitem [{\citenamefont {Yao}\ \emph {et~al.}(2023)\citenamefont {Yao}, \citenamefont {Gaudet}, \citenamefont {Verma}, \citenamefont {Graf}, \citenamefont {Yang}, \citenamefont {Bahrami}, \citenamefont {Zhang}, \citenamefont {Aczel}, \citenamefont {Subedi}, \citenamefont {Torchinsky} \emph {et~al.}}]{yao2023large}%
  \BibitemOpen
  \bibfield  {author} {\bibinfo {author} {\bibfnamefont {X.}~\bibnamefont {Yao}}, \bibinfo {author} {\bibfnamefont {J.}~\bibnamefont {Gaudet}}, \bibinfo {author} {\bibfnamefont {R.}~\bibnamefont {Verma}}, \bibinfo {author} {\bibfnamefont {D.~E.}\ \bibnamefont {Graf}}, \bibinfo {author} {\bibfnamefont {H.~Y.}\ \bibnamefont {Yang}}, \bibinfo {author} {\bibfnamefont {F.}~\bibnamefont {Bahrami}}, \bibinfo {author} {\bibfnamefont {R.}~\bibnamefont {Zhang}}, \bibinfo {author} {\bibfnamefont {A.~A.}\ \bibnamefont {Aczel}}, \bibinfo {author} {\bibfnamefont {S.}~\bibnamefont {Subedi}}, \bibinfo {author} {\bibfnamefont {D.~H.}\ \bibnamefont {Torchinsky}}, \emph {et~al.},\ }\bibfield  {title} {\bibinfo {title} {Large topological {H}all effect and spiral magnetic order in the {W}eyl semimetal \ch{SmAlSi}},\ }{\bibfield  {journal} {\bibinfo  {journal} {Physical Review X}\ }\textbf {\bibinfo {volume} {13}},\ \bibinfo {pages} {011035} (\bibinfo {year} {2023})}\BibitemShut {NoStop}%
\bibitem [{\citenamefont {Yamada}\ \emph {et~al.}(2024)\citenamefont {Yamada}, \citenamefont {Nomoto}, \citenamefont {Miyake}, \citenamefont {Terakawa}, \citenamefont {Kikkawa}, \citenamefont {Arita}, \citenamefont {Tokunaga}, \citenamefont {Taguchi}, \citenamefont {Tokura},\ and\ \citenamefont {Hirschberger}}]{yamada2024nernst}%
  \BibitemOpen
  \bibfield  {author} {\bibinfo {author} {\bibfnamefont {R.}~\bibnamefont {Yamada}}, \bibinfo {author} {\bibfnamefont {T.}~\bibnamefont {Nomoto}}, \bibinfo {author} {\bibfnamefont {A.}~\bibnamefont {Miyake}}, \bibinfo {author} {\bibfnamefont {T.}~\bibnamefont {Terakawa}}, \bibinfo {author} {\bibfnamefont {A.}~\bibnamefont {Kikkawa}}, \bibinfo {author} {\bibfnamefont {R.}~\bibnamefont {Arita}}, \bibinfo {author} {\bibfnamefont {M.}~\bibnamefont {Tokunaga}}, \bibinfo {author} {\bibfnamefont {Y.}~\bibnamefont {Taguchi}}, \bibinfo {author} {\bibfnamefont {Y.}~\bibnamefont {Tokura}},\ and\ \bibinfo {author} {\bibfnamefont {M.}~\bibnamefont {Hirschberger}},\ }\bibfield  {title} {\bibinfo {title} {Nernst effect of high-mobility Weyl electrons in \ch{NdAlSi} enhanced by a Fermi surface nesting instability},\ }{\bibfield  {journal} {\bibinfo  {journal} {Physical Review X}\ }\textbf {\bibinfo {volume} {14}},\ \bibinfo {pages} {021012} (\bibinfo {year} {2024})}\BibitemShut {NoStop}%
\bibitem [{\citenamefont {Wang}\ \emph {et~al.}(2022)\citenamefont {Wang}, \citenamefont {Dong}, \citenamefont {Guo}, \citenamefont {Lv}, \citenamefont {Huang}, \citenamefont {Xiang}, \citenamefont {Ren}, \citenamefont {Wang}, \citenamefont {Sun}, \citenamefont {Li} \emph {et~al.}}]{wang2022ndalsi}%
  \BibitemOpen
  \bibfield  {author} {\bibinfo {author} {\bibfnamefont {J.-F.}\ \bibnamefont {Wang}}, \bibinfo {author} {\bibfnamefont {Q.-X.}\ \bibnamefont {Dong}}, \bibinfo {author} {\bibfnamefont {Z.-P.}\ \bibnamefont {Guo}}, \bibinfo {author} {\bibfnamefont {M.}~\bibnamefont {Lv}}, \bibinfo {author} {\bibfnamefont {Y.-F.}\ \bibnamefont {Huang}}, \bibinfo {author} {\bibfnamefont {J.-S.}\ \bibnamefont {Xiang}}, \bibinfo {author} {\bibfnamefont {Z.-A.}\ \bibnamefont {Ren}}, \bibinfo {author} {\bibfnamefont {Z.-J.}\ \bibnamefont {Wang}}, \bibinfo {author} {\bibfnamefont {P.-J.}\ \bibnamefont {Sun}}, \bibinfo {author} {\bibfnamefont {G.}~\bibnamefont {Li}}, \emph {et~al.},\ }\bibfield  {title} {\bibinfo {title} {\ch{NdAlSi}: A magnetic Weyl semimetal candidate with rich magnetic phases and atypical transport properties},\ }{\bibfield  {journal} {\bibinfo  {journal} {Physical Review B}\ }\textbf {\bibinfo {volume} {105}},\ \bibinfo {pages} {144435} (\bibinfo {year} {2022})}\BibitemShut {NoStop}%
\bibitem [{\citenamefont {Wang}\ \emph {et~al.}(2023)\citenamefont {Wang}, \citenamefont {Dong}, \citenamefont {Huang}, \citenamefont {Wang}, \citenamefont {Guo}, \citenamefont {Wang}, \citenamefont {Ren}, \citenamefont {Li}, \citenamefont {Sun}, \citenamefont {Dai} \emph {et~al.}}]{wang2023quantum}%
  \BibitemOpen
  \bibfield  {author} {\bibinfo {author} {\bibfnamefont {J.-F.}\ \bibnamefont {Wang}}, \bibinfo {author} {\bibfnamefont {Q.-X.}\ \bibnamefont {Dong}}, \bibinfo {author} {\bibfnamefont {Y.-F.}\ \bibnamefont {Huang}}, \bibinfo {author} {\bibfnamefont {Z.-S.}\ \bibnamefont {Wang}}, \bibinfo {author} {\bibfnamefont {Z.-P.}\ \bibnamefont {Guo}}, \bibinfo {author} {\bibfnamefont {Z.-J.}\ \bibnamefont {Wang}}, \bibinfo {author} {\bibfnamefont {Z.-A.}\ \bibnamefont {Ren}}, \bibinfo {author} {\bibfnamefont {G.}~\bibnamefont {Li}}, \bibinfo {author} {\bibfnamefont {P.-J.}\ \bibnamefont {Sun}}, \bibinfo {author} {\bibfnamefont {X.}~\bibnamefont {Dai}}, \emph {et~al.},\ }\bibfield  {title} {\bibinfo {title} {Quantum oscillations in the magnetic Weyl semimetal \ch{NdAlSi} arising from strong Weyl fermion--4f electron exchange interaction},\ } {\bibfield  {journal} {\bibinfo  {journal} {Physical Review B}\ }\textbf {\bibinfo {volume} {108}},\ \bibinfo {pages} {024423} (\bibinfo {year} {2023})}\BibitemShut
  {NoStop}%
\bibitem [{\citenamefont {Piva}\ \emph {et~al.}(2023)\citenamefont {Piva}, \citenamefont {Souza}, \citenamefont {Lombardi}, \citenamefont {Pakuszewski}, \citenamefont {Adriano}, \citenamefont {Pagliuso},\ and\ \citenamefont {Nicklas}}]{piva2023topological}%
  \BibitemOpen
  \bibfield  {author} {\bibinfo {author} {\bibfnamefont {M.}~\bibnamefont {Piva}}, \bibinfo {author} {\bibfnamefont {J.}~\bibnamefont {Souza}}, \bibinfo {author} {\bibfnamefont {G.}~\bibnamefont {Lombardi}}, \bibinfo {author} {\bibfnamefont {K.}~\bibnamefont {Pakuszewski}}, \bibinfo {author} {\bibfnamefont {C.}~\bibnamefont {Adriano}}, \bibinfo {author} {\bibfnamefont {P.}~\bibnamefont {Pagliuso}},\ and\ \bibinfo {author} {\bibfnamefont {M.}~\bibnamefont {Nicklas}},\ }\bibfield  {title} {\bibinfo {title} {Topological Hall effect in \ch{CeAlGe}},\ }{\bibfield  {journal} {\bibinfo  {journal} {Physical Review Materials}\ }\textbf {\bibinfo {volume} {7}},\ \bibinfo {pages} {074204} (\bibinfo {year} {2023})}\BibitemShut {NoStop}%
\bibitem [{\citenamefont {Ram}\ \emph {et~al.}(2023{\natexlab{a}})\citenamefont {Ram}, \citenamefont {Malick}, \citenamefont {Hossain},\ and\ \citenamefont {Kaczorowski}}]{ram2023magnetic}%
  \BibitemOpen
  \bibfield  {author} {\bibinfo {author} {\bibfnamefont {D.}~\bibnamefont {Ram}}, \bibinfo {author} {\bibfnamefont {S.}~\bibnamefont {Malick}}, \bibinfo {author} {\bibfnamefont {Z.}~\bibnamefont {Hossain}},\ and\ \bibinfo {author} {\bibfnamefont {D.}~\bibnamefont {Kaczorowski}},\ }\bibfield  {title} {\bibinfo {title} {Magnetic, thermodynamic, and magnetotransport properties of \ch{CeGaGe} and \ch{PrGaGe} single crystals},\ } {\bibfield  {journal} {\bibinfo  {journal} {Physical Review B}\ }\textbf {\bibinfo {volume} {108}},\ \bibinfo {pages} {024428} (\bibinfo {year} {2023}{\natexlab{a}})}\BibitemShut {NoStop}%
\bibitem [{\citenamefont {Gong}\ \emph {et~al.}(2024)\citenamefont {Gong}, \citenamefont {Wang}, \citenamefont {Han}, \citenamefont {Zeng}, \citenamefont {Ma}, \citenamefont {Wang}, \citenamefont {Lin}, \citenamefont {Wang},\ and\ \citenamefont {Xia}}]{gong2024anomalous}%
  \BibitemOpen
  \bibfield  {author} {\bibinfo {author} {\bibfnamefont {J.}~\bibnamefont {Gong}}, \bibinfo {author} {\bibfnamefont {H.}~\bibnamefont {Wang}}, \bibinfo {author} {\bibfnamefont {K.}~\bibnamefont {Han}}, \bibinfo {author} {\bibfnamefont {X.-Y.}\ \bibnamefont {Zeng}}, \bibinfo {author} {\bibfnamefont {X.-P.}\ \bibnamefont {Ma}}, \bibinfo {author} {\bibfnamefont {Y.-T.}\ \bibnamefont {Wang}}, \bibinfo {author} {\bibfnamefont {J.-F.}\ \bibnamefont {Lin}}, \bibinfo {author} {\bibfnamefont {X.-Y.}\ \bibnamefont {Wang}},\ and\ \bibinfo {author} {\bibfnamefont {T.-L.}\ \bibnamefont {Xia}},\ }\bibfield  {title} {\bibinfo {title} {Anomalous Hall effect in an antiferromagnetic \ch{CeGaSi} single crystal},\ }{\bibfield  {journal} {\bibinfo  {journal} {Physical Review B}\ }\textbf {\bibinfo {volume} {109}},\ \bibinfo {pages} {024434} (\bibinfo {year} {2024})}\BibitemShut {NoStop}%
\bibitem [{\citenamefont {Zhang}\ \emph {et~al.}(2024)\citenamefont {Zhang}, \citenamefont {Dong}, \citenamefont {Bai}, \citenamefont {Liu}, \citenamefont {Cheng}, \citenamefont {Li}, \citenamefont {Liu}, \citenamefont {Sun}, \citenamefont {Huang}, \citenamefont {Ren} \emph {et~al.}}]{zhang2024magnetism}%
  \BibitemOpen
  \bibfield  {author} {\bibinfo {author} {\bibfnamefont {L.-B.}\ \bibnamefont {Zhang}}, \bibinfo {author} {\bibfnamefont {Q.-X.}\ \bibnamefont {Dong}}, \bibinfo {author} {\bibfnamefont {J.-L.}\ \bibnamefont {Bai}}, \bibinfo {author} {\bibfnamefont {Q.-Y.}\ \bibnamefont {Liu}}, \bibinfo {author} {\bibfnamefont {J.-W.}\ \bibnamefont {Cheng}}, \bibinfo {author} {\bibfnamefont {C.-D.}\ \bibnamefont {Li}}, \bibinfo {author} {\bibfnamefont {P.-Y.}\ \bibnamefont {Liu}}, \bibinfo {author} {\bibfnamefont {Y.-R.}\ \bibnamefont {Sun}}, \bibinfo {author} {\bibfnamefont {Y.}~\bibnamefont {Huang}}, \bibinfo {author} {\bibfnamefont {Z.-A.}\ \bibnamefont {Ren}}, \emph {et~al.},\ }\bibfield  {title} {\bibinfo {title} {Magnetism, heat capacity, magnetocaloric effect, and magneto-transport properties of heavy fermion antiferromagnet \ch{CeGaSi}},\ }{\bibfield  {journal} {\bibinfo  {journal} {Chinese Physics B}\ }\textbf {\bibinfo {volume} {33}},\ \bibinfo {pages} {067101} (\bibinfo {year} {2024})}\BibitemShut
  {NoStop}%
\bibitem [{\citenamefont {Nagaosa}\ \emph {et~al.}(2010)\citenamefont {Nagaosa}, \citenamefont {Sinova}, \citenamefont {Onoda}, \citenamefont {MacDonald},\ and\ \citenamefont {Ong}}]{nagaosa2010anomalous}%
  \BibitemOpen
  \bibfield  {author} {\bibinfo {author} {\bibfnamefont {N.}~\bibnamefont {Nagaosa}}, \bibinfo {author} {\bibfnamefont {J.}~\bibnamefont {Sinova}}, \bibinfo {author} {\bibfnamefont {S.}~\bibnamefont {Onoda}}, \bibinfo {author} {\bibfnamefont {A.~H.}\ \bibnamefont {MacDonald}},\ and\ \bibinfo {author} {\bibfnamefont {N.~P.}\ \bibnamefont {Ong}},\ }\bibfield  {title} {\bibinfo {title} {Anomalous {H}all effect},\ } {\bibfield  {journal} {\bibinfo  {journal} {Reviews of Modern Physics}\ }\textbf {\bibinfo {volume} {82}},\ \bibinfo {pages} {1539} (\bibinfo {year} {2010})}\BibitemShut {NoStop}%
\bibitem [{\citenamefont {K{\"u}bler}\ and\ \citenamefont {Felser}(2014)}]{kubler2014non}%
  \BibitemOpen
  \bibfield  {author} {\bibinfo {author} {\bibfnamefont {J.}~\bibnamefont {K{\"u}bler}}\ and\ \bibinfo {author} {\bibfnamefont {C.}~\bibnamefont {Felser}},\ }\bibfield  {title} {\bibinfo {title} {Non-collinear antiferromagnets and the anomalous {H}all effect},\ }{\bibfield  {journal} {\bibinfo  {journal} {Europhysics Letters}\ }\textbf {\bibinfo {volume} {108}},\ \bibinfo {pages} {67001} (\bibinfo {year} {2014})}\BibitemShut {NoStop}%
\bibitem [{\citenamefont 
{{\u{S}}mejkal}\ \emph {et~al.}(2015)\citenamefont {{\u{S}}mejkal}, \citenamefont {MacDonald},\citenamefont {Sinova},\citenamefont {Nakatsuji},\ and\ \citenamefont {Jungwirth}}]{aheafm2022libor}%
  \BibitemOpen
  \bibfield  {author} {\bibinfo {author} {\bibfnamefont {L.}~\bibnamefont {{\u{S}}mejkal}}, \bibinfo {author} {\bibfnamefont {A.~H.}~\bibnamefont {Macdonald}},\bibinfo {author} {\bibfnamefont {J.}~\bibnamefont {Sinova}},\bibinfo {author} {\bibfnamefont {S.}~\bibnamefont {Nakatsuji}},\ and\ \bibinfo {author} {\bibfnamefont {T.}~\bibnamefont {Jungwirth}},\ }\bibfield  {title} {\bibinfo {title} {Anomalous Hall antiferromagnets},\ }{\bibfield  {journal} {\bibinfo  {journal} {Nature Review Materials}\ }\textbf {\bibinfo {volume} {7}},\ \bibinfo {pages} {482-496} (\bibinfo {year} {2022})}\BibitemShut {NoStop}%
\bibitem [{\citenamefont 
{Nakatsuji}\ \emph {et~al.}(2015)\citenamefont {Nakatsuji}, \citenamefont {Kiyohara},\ and\ \citenamefont {Higo}}]{nakatsuji2015large}%
  \BibitemOpen
  \bibfield  {author} {\bibinfo {author} {\bibfnamefont {S.}~\bibnamefont {Nakatsuji}}, \bibinfo {author} {\bibfnamefont {N.}~\bibnamefont {Kiyohara}},\ and\ \bibinfo {author} {\bibfnamefont {T.}~\bibnamefont {Higo}},\ }\bibfield  {title} {\bibinfo {title} {Large anomalous {H}all effect in a non-collinear antiferromagnet at room temperature},\ }{\bibfield  {journal} {\bibinfo  {journal} {Nature}\ }\textbf {\bibinfo {volume} {527}},\ \bibinfo {pages} {212} (\bibinfo {year} {2015})}\BibitemShut {NoStop}%
\bibitem [{\citenamefont {Nayak}\ \emph {et~al.}(2016)\citenamefont {Nayak}, \citenamefont {Fischer}, \citenamefont {Sun}, \citenamefont {Yan}, \citenamefont {Karel}, \citenamefont {Komarek}, \citenamefont {Shekhar}, \citenamefont {Kumar}, \citenamefont {Schnelle}, \citenamefont {K{\"u}bler} \emph {et~al.}}]{nayak2016large}%
  \BibitemOpen
  \bibfield  {author} {\bibinfo {author} {\bibfnamefont {A.~K.}\ \bibnamefont {Nayak}}, \bibinfo {author} {\bibfnamefont {J.~E.}\ \bibnamefont {Fischer}}, \bibinfo {author} {\bibfnamefont {Y.}~\bibnamefont {Sun}}, \bibinfo {author} {\bibfnamefont {B.}~\bibnamefont {Yan}}, \bibinfo {author} {\bibfnamefont {J.}~\bibnamefont {Karel}}, \bibinfo {author} {\bibfnamefont {A.~C.}\ \bibnamefont {Komarek}}, \bibinfo {author} {\bibfnamefont {C.}~\bibnamefont {Shekhar}}, \bibinfo {author} {\bibfnamefont {N.}~\bibnamefont {Kumar}}, \bibinfo {author} {\bibfnamefont {W.}~\bibnamefont {Schnelle}}, \bibinfo {author} {\bibfnamefont {J.}~\bibnamefont {K{\"u}bler}}, \emph {et~al.},\ }\bibfield  {title} {\bibinfo {title} {Large anomalous {H}all effect driven by a nonvanishing {B}erry curvature in the noncolinear antiferromagnet \ch{Mn3Ge}},\ }{\bibfield  {journal} {\bibinfo  {journal} {Science Advances}\ }\textbf {\bibinfo {volume} {2}},\ \bibinfo {pages} {e1501870} (\bibinfo {year} {2016})}\BibitemShut {NoStop}%
   \bibitem [{\citenamefont {Zhao}\ \emph {et~al.}(2020)\citenamefont {Zhao}, \citenamefont {Deng}, \citenamefont {Chen}, \citenamefont {Ross}, \citenamefont {Pet{\v{r}}{\'{i}}{\v{c}}ek}, \citenamefont {Noky}, \citenamefont {G{\"{u}}nther}, \citenamefont {Russina},\citenamefont {Hutanu},\ and\ \citenamefont {Gegenwart}}]{zhao2020ice}%
  \BibitemOpen
  \bibfield  {author} {\bibinfo {author} {\bibfnamefont {K.}~\bibnamefont {Zhao}}, \bibinfo {author} {\bibfnamefont {H.}~\bibnamefont {Deng}}, \bibinfo {author} {\bibfnamefont {H.}~\bibnamefont {Chen}}, \bibinfo {author} {\bibfnamefont {K.~A.}~\bibnamefont {Ross}}, \bibinfo {author} {\bibfnamefont {V.}~\bibnamefont {Pet{\v{r}}{\'{i}}{\v{c}}ek}}, \bibinfo {author} {\bibfnamefont {G.}~\bibnamefont {G{\"{u}}nther}}, \bibinfo {author} {\bibfnamefont {M.}\ \bibnamefont {Russina}}, \bibinfo {author} {\bibfnamefont {V.}~\bibnamefont {Hutanu}},\ and\ \bibinfo {author} {\bibfnamefont {P.}~\bibnamefont {Gegenwart}},\ }\bibfield  {title} {\bibinfo {title} {Realization of the kagome spin ice state in a frustrated intermetallic compound},\ } {\bibfield  {journal} {\bibinfo  {journal} {Science}\ }\textbf {\bibinfo {volume} {367}},\ \bibinfo {pages} {1218-1223} (\bibinfo {year} {2020})}\BibitemShut {NoStop}%
  \bibitem [{\citenamefont {Zhao}\ \emph {et~al.}(2020)\citenamefont {Zhao}, \citenamefont {Tokiwa}, \citenamefont {Chen},\ and\ \citenamefont {Gegenwart}}]{zhao2024ahe}%
  \BibitemOpen
  \bibfield  {author} {\bibinfo {author} {\bibfnamefont {K.}~\bibnamefont {Zhao}}, \bibinfo {author} {\bibfnamefont {Y.}~\bibnamefont {Tokiwa}}, \bibinfo {author} {\bibfnamefont {H.}~\bibnamefont {Chen}},\ and\ \bibinfo {author} {\bibfnamefont {P.}~\bibnamefont {Gegenwart}},\ }\bibfield  {title} {\bibinfo {title} {Discrete degeneracies distinguished by the anomalous Hall effect in a metallic kagome ice compound},\ } {\bibfield  {journal} {\bibinfo  {journal} {Nature Physcs}\ }\textbf {\bibinfo {volume} {20}},\ \bibinfo {pages} {442–449} (\bibinfo {year} {2024})}\BibitemShut {NoStop}%
  \bibitem [{\citenamefont {Roychowdhury}\ \emph {et~al.}(2024)\citenamefont {Roychowdhury}, \citenamefont {Samanta}, \citenamefont {Singh}, \citenamefont {Schnelle}, \citenamefont {Zhang}, \citenamefont {Noky}, \citenamefont {Vergniory}, \citenamefont {Shekhar},\ and\ \citenamefont {Felser}}]{roychowdhury2024enhancement}%
  \BibitemOpen
  \bibfield  {author} {\bibinfo {author} {\bibfnamefont {S.}~\bibnamefont {Roychowdhury}}, \bibinfo {author} {\bibfnamefont {K.}~\bibnamefont {Samanta}}, \bibinfo {author} {\bibfnamefont {S.}~\bibnamefont {Singh}}, \bibinfo {author} {\bibfnamefont {W.}~\bibnamefont {Schnelle}}, \bibinfo {author} {\bibfnamefont {Y.}~\bibnamefont {Zhang}}, \bibinfo {author} {\bibfnamefont {J.}~\bibnamefont {Noky}}, \bibinfo {author} {\bibfnamefont {M.~G.}\ \bibnamefont {Vergniory}}, \bibinfo {author} {\bibfnamefont {C.}~\bibnamefont {Shekhar}},\ and\ \bibinfo {author} {\bibfnamefont {C.}~\bibnamefont {Felser}},\ }\bibfield  {title} {\bibinfo {title} {Enhancement of the anomalous {H}all effect by distorting the {K}agom{\'e} lattice in an antiferromagnetic material},\ } {\bibfield  {journal} {\bibinfo  {journal} {Proceedings of the National Academy of Sciences}\ }\textbf {\bibinfo {volume} {121}},\ \bibinfo {pages} {e2401970121} (\bibinfo {year} {2024})}\BibitemShut {NoStop}%
\bibitem [{sup()}]{supply}%
  \BibitemOpen
  {\bibinfo {title} {See the {S}upplemental {M}aterial for the details of sample characterization and additional {H}all resistivity analysis.}}\BibitemShut {Stop}%
\bibitem [{\citenamefont {Das}\ \emph {et~al.}(2012)\citenamefont {Das}, \citenamefont {Kumar}, \citenamefont {Kulkarni}, \citenamefont {Dhar},\ and\ \citenamefont {Thamizhavel}}]{das2012anisotropic}%
  \BibitemOpen
  \bibfield  {author} {\bibinfo {author} {\bibfnamefont {P.~K.}\ \bibnamefont {Das}}, \bibinfo {author} {\bibfnamefont {N.}~\bibnamefont {Kumar}}, \bibinfo {author} {\bibfnamefont {R.}~\bibnamefont {Kulkarni}}, \bibinfo {author} {\bibfnamefont {S.}~\bibnamefont {Dhar}},\ and\ \bibinfo {author} {\bibfnamefont {A.}~\bibnamefont {Thamizhavel}},\ }\bibfield  {title} {\bibinfo {title} {Anisotropic magnetic properties and superzone gap formation in \ch{CeGe} single crystal},\ } {\bibfield  {journal} {\bibinfo  {journal} {Journal of Physics: Condensed Matter}\ }\textbf {\bibinfo {volume} {24}},\ \bibinfo {pages} {146003} (\bibinfo {year} {2012})}\BibitemShut {NoStop}%
\bibitem [{\citenamefont {Gupta}\ \emph {et~al.}(2015)\citenamefont {Gupta}, \citenamefont {Suresh}, \citenamefont {Das}, \citenamefont {Nigam},\ and\ \citenamefont {Hoser}}]{gupta}%
  \BibitemOpen
  \bibfield  {author} {\bibinfo {author} {\bibfnamefont {S.}~\bibnamefont {Gupta}}, \bibinfo {author} {\bibfnamefont {K.~G.}\ \bibnamefont {Suresh}}, \bibinfo {author} {\bibfnamefont {A.}~\bibnamefont {Das}}, \bibinfo {author} {\bibfnamefont {A.~K.}\ \bibnamefont {Nigam}},\ and\ \bibinfo {author} {\bibfnamefont {A.}~\bibnamefont {Hoser}},\ }\bibfield  {title} {\bibinfo {title} {{Effects of antiferro-ferromagnetic phase coexistence and spin fluctuations on the magnetic and related properties of \ch{NdCuSi}}},\ }{\bibfield  {journal} {\bibinfo  {journal} {APL Materials}\ }\textbf {\bibinfo {volume} {3}},\ \bibinfo {pages} {066102} (\bibinfo {year} {2015})}\BibitemShut {NoStop}%
\bibitem [{\citenamefont {Hodovanets}\ \emph {et~al.}(2018)\citenamefont {Hodovanets}, \citenamefont {Eckberg}, \citenamefont {Zavalij}, \citenamefont {Kim}, \citenamefont {Lin}, \citenamefont {Zic}, \citenamefont {Campbell}, \citenamefont {Higgins},\ and\ \citenamefont {Paglione}}]{hodovanets2018single}%
  \BibitemOpen
  \bibfield  {author} {\bibinfo {author} {\bibfnamefont {H.}~\bibnamefont {Hodovanets}}, \bibinfo {author} {\bibfnamefont {C.}~\bibnamefont {Eckberg}}, \bibinfo {author} {\bibfnamefont {P.}~\bibnamefont {Zavalij}}, \bibinfo {author} {\bibfnamefont {H.}~\bibnamefont {Kim}}, \bibinfo {author} {\bibfnamefont {W.-C.}\ \bibnamefont {Lin}}, \bibinfo {author} {\bibfnamefont {M.}~\bibnamefont {Zic}}, \bibinfo {author} {\bibfnamefont {D.}~\bibnamefont {Campbell}}, \bibinfo {author} {\bibfnamefont {J.}~\bibnamefont {Higgins}},\ and\ \bibinfo {author} {\bibfnamefont {J.}~\bibnamefont {Paglione}},\ }\bibfield  {title} {\bibinfo {title} {Single-crystal investigation of the proposed type-{II} {W}eyl semimetal \ch{CeAlGe}},\ }{\bibfield  {journal} {\bibinfo  {journal} {Physical Review B}\ }\textbf {\bibinfo {volume} {98}},\ \bibinfo {pages} {245132} (\bibinfo {year} {2018})}\BibitemShut {NoStop}%
\bibitem [{\citenamefont {Arantes}\ \emph {et~al.}(2018)\citenamefont {Arantes}, \citenamefont {Aristiz{\'a}bal-Giraldo}, \citenamefont {Masunaga}, \citenamefont {Costa}, \citenamefont {Ferreira}, \citenamefont {Takabatake}, \citenamefont {Mendonca-Ferreira}, \citenamefont {Ribeiro},\ and\ \citenamefont {Avila}}]{arantes2018structure}%
  \BibitemOpen
  \bibfield  {author} {\bibinfo {author} {\bibfnamefont {F.~R.}\ \bibnamefont {Arantes}}, \bibinfo {author} {\bibfnamefont {D.}~\bibnamefont {Aristiz{\'a}bal-Giraldo}}, \bibinfo {author} {\bibfnamefont {S.~H.}\ \bibnamefont {Masunaga}}, \bibinfo {author} {\bibfnamefont {F.~N.}\ \bibnamefont {Costa}}, \bibinfo {author} {\bibfnamefont {F.~F.}\ \bibnamefont {Ferreira}}, \bibinfo {author} {\bibfnamefont {T.}~\bibnamefont {Takabatake}}, \bibinfo {author} {\bibfnamefont {L.}~\bibnamefont {Mendonca-Ferreira}}, \bibinfo {author} {\bibfnamefont {R.~A.}\ \bibnamefont {Ribeiro}},\ and\ \bibinfo {author} {\bibfnamefont {M.~A.}\ \bibnamefont {Avila}},\ }\bibfield  {title} {\bibinfo {title} {Structure, magnetism, and transport of single-crystalline \ch{RNiSi3} ({R} = {Y}, {G}d-{T}m, {L}u)},\ } {\bibfield  {journal} {\bibinfo  {journal} {Physical Review Materials}\ }\textbf {\bibinfo {volume} {2}},\ \bibinfo {pages} {044402} (\bibinfo {year} {2018})}\BibitemShut {NoStop}%
\bibitem [{\citenamefont {Muthuselvam}\ \emph {et~al.}(2019)\citenamefont {Muthuselvam}, \citenamefont {Nehru}, \citenamefont {Babu}, \citenamefont {Saranya}, \citenamefont {Kaul}, \citenamefont {Chen}, \citenamefont {Chen}, \citenamefont {Liu}, \citenamefont {Guo}, \citenamefont {Xiu} \emph {et~al.}}]{muthuselvam2019gd2te3}%
  \BibitemOpen
  \bibfield  {author} {\bibinfo {author} {\bibfnamefont {I.~P.}\ \bibnamefont {Muthuselvam}}, \bibinfo {author} {\bibfnamefont {R.}~\bibnamefont {Nehru}}, \bibinfo {author} {\bibfnamefont {K.~R.}\ \bibnamefont {Babu}}, \bibinfo {author} {\bibfnamefont {K.}~\bibnamefont {Saranya}}, \bibinfo {author} {\bibfnamefont {S.}~\bibnamefont {Kaul}}, \bibinfo {author} {\bibfnamefont {S.-M.}\ \bibnamefont {Chen}}, \bibinfo {author} {\bibfnamefont {W.-T.}\ \bibnamefont {Chen}}, \bibinfo {author} {\bibfnamefont {Y.}~\bibnamefont {Liu}}, \bibinfo {author} {\bibfnamefont {G.-Y.}\ \bibnamefont {Guo}}, \bibinfo {author} {\bibfnamefont {F.}~\bibnamefont {Xiu}}, \emph {et~al.},\ }\bibfield  {title} {\bibinfo {title} {\ch{Gd2Te3}: an antiferromagnetic semimetal},\ }{\bibfield  {journal} {\bibinfo  {journal} {Journal of Physics: Condensed Matter}\ }\textbf {\bibinfo {volume} {31}},\ \bibinfo {pages} {285802} (\bibinfo {year} {2019})}\BibitemShut {NoStop}%
\bibitem [{\citenamefont {Ram}\ \emph {et~al.}(2023{\natexlab{b}})\citenamefont {Ram}, \citenamefont {Singh}, \citenamefont {Hooda}, \citenamefont {Singh}, \citenamefont {Kanchana}, \citenamefont {Kaczorowski},\ and\ \citenamefont {Hossain}}]{ram2023multiple}%
  \BibitemOpen
  \bibfield  {author} {\bibinfo {author} {\bibfnamefont {D.}~\bibnamefont {Ram}}, \bibinfo {author} {\bibfnamefont {J.}~\bibnamefont {Singh}}, \bibinfo {author} {\bibfnamefont {M.}~\bibnamefont {Hooda}}, \bibinfo {author} {\bibfnamefont {K.}~\bibnamefont {Singh}}, \bibinfo {author} {\bibfnamefont {V.}~\bibnamefont {Kanchana}}, \bibinfo {author} {\bibfnamefont {D.}~\bibnamefont {Kaczorowski}},\ and\ \bibinfo {author} {\bibfnamefont {Z.}~\bibnamefont {Hossain}},\ }\bibfield  {title} {\bibinfo {title} {Multiple magnetic transitions, metamagnetism, and large magnetoresistance in \ch{GdAuGe} single crystals},\ } {\bibfield  {journal} {\bibinfo  {journal} {Physical Review B}\ }\textbf {\bibinfo {volume} {108}},\ \bibinfo {pages} {235107} (\bibinfo {year} {2023}{\natexlab{b}})}\BibitemShut {NoStop}%
\bibitem [{\citenamefont {Zhang}\ \emph {et~al.}(1994)\citenamefont {Zhang}, \citenamefont {Gignoux}, \citenamefont {Schmitt}, \citenamefont {Franse}, \citenamefont {Kayzel}, \citenamefont {Kim-Ngan},\ and\ \citenamefont {Radwanski}}]{zhang1994crystalline}%
  \BibitemOpen
  \bibfield  {author} {\bibinfo {author} {\bibfnamefont {F.}~\bibnamefont {Zhang}}, \bibinfo {author} {\bibfnamefont {D.}~\bibnamefont {Gignoux}}, \bibinfo {author} {\bibfnamefont {D.}~\bibnamefont {Schmitt}}, \bibinfo {author} {\bibfnamefont {J.}~\bibnamefont {Franse}}, \bibinfo {author} {\bibfnamefont {F.}~\bibnamefont {Kayzel}}, \bibinfo {author} {\bibfnamefont {N.}~\bibnamefont {Kim-Ngan}},\ and\ \bibinfo {author} {\bibfnamefont {R.}~\bibnamefont {Radwanski}},\ }\bibfield  {title} {\bibinfo {title} {Crystalline electric field and high field magnetization in \ch{ErNi5} single crystal},\ }{\bibfield  {journal} {\bibinfo  {journal} {Journal of Magnetism and Magnetic Materials}\ }\textbf {\bibinfo {volume} {130}},\ \bibinfo {pages} {108} (\bibinfo {year} {1994})}\BibitemShut {NoStop}%
\bibitem [{\citenamefont {Gopal}(2012)}]{gopal2012specific}%
  \BibitemOpen
  \bibfield  {author} {\bibinfo {author} {\bibfnamefont {E.}~\bibnamefont {Gopal}},\ }{\emph {\bibinfo {title} {Specific heats at low temperatures}}}\ (\bibinfo  {publisher} {Springer Science \& Business Media},\ \bibinfo {year} {2012})\BibitemShut {NoStop}%
\bibitem [{\citenamefont {Campoy}\ \emph {et~al.}(2006)\citenamefont {Campoy}, \citenamefont {Plaza}, \citenamefont {Coelho},\ and\ \citenamefont {Gama}}]{campoy2006magnetoresistivity}%
  \BibitemOpen
  \bibfield  {author} {\bibinfo {author} {\bibfnamefont {J.}~\bibnamefont {Campoy}}, \bibinfo {author} {\bibfnamefont {E.}~\bibnamefont {Plaza}}, \bibinfo {author} {\bibfnamefont {A.}~\bibnamefont {Coelho}},\ and\ \bibinfo {author} {\bibfnamefont {S.}~\bibnamefont {Gama}},\ }\bibfield  {title} {\bibinfo {title} {Magnetoresistivity as a probe to the field-induced change of magnetic entropy in \ch{RAl2} compounds ({R} = {P}r, {N}d, {T}b, {D}y, {H}o, {E}r)},\ } {\bibfield  {journal} {\bibinfo  {journal} {Physical Review B}\ }\textbf {\bibinfo {volume} {74}},\ \bibinfo {pages} {134410} (\bibinfo {year} {2006})}\BibitemShut {NoStop}%
\bibitem [{\citenamefont {Szytu{\l}a}\ \emph {et~al.}(2007)\citenamefont {Szytu{\l}a}, \citenamefont {Kaczorowski}, \citenamefont {Nenkov} \emph {et~al.}}]{szytula2007electronic}%
  \BibitemOpen
  \bibfield  {author} {\bibinfo {author} {\bibfnamefont {A.}~\bibnamefont {Szytu{\l}a}}, \bibinfo {author} {\bibfnamefont {D.}~\bibnamefont {Kaczorowski}}, \bibinfo {author} {\bibfnamefont {K.}~\bibnamefont {Nenkov}}, \emph {et~al.},\ }\bibfield  {title} {\bibinfo {title} {Electronic structure and magnetism of \ch{RPdIn} compounds ({R} = {L}a, {C}e, {P}r, {N}d)},\ }{\bibfield  {journal} {\bibinfo  {journal} {Solid State Communications}\ }\textbf {\bibinfo {volume} {142}},\ \bibinfo {pages} {556} (\bibinfo {year} {2007})}\BibitemShut {NoStop}%
\bibitem [{\citenamefont {Walter}(2014)}]{waltercef}%
  \BibitemOpen
  \bibfield  {author} {\bibinfo {author} {\bibfnamefont {U.}~\bibnamefont {Walter}},\ }\bibfield  {title} {\bibinfo {title} {Treating crystal field parameters in lower than cubic symmetries},\ }{\bibfield  {journal} {\bibinfo  {journal} {Journal of physics and chemistry of solids}\ }\textbf {\bibinfo {volume} {45}},\ \bibinfo {pages} {401-408} (\bibinfo {year} {1984})}\BibitemShut {NoStop}%
\bibitem [{\citenamefont {Allenspach}\ \emph {et~al.}(1994)\citenamefont {Allenspach}, \citenamefont {Mesot}, \citenamefont {Staub}, \citenamefont {Guillaume}, \citenamefont {Furrer}, \citenamefont {Yoo}, \citenamefont {Kramer}, \citenamefont {McCallum}, \citenamefont {Maletta}, \citenamefont {Blank} \emph {et~al.}}]{allenspach1994magnetic}%
  \BibitemOpen
  \bibfield  {author} {\bibinfo {author} {\bibfnamefont {P.}~\bibnamefont {Allenspach}}, \bibinfo {author} {\bibfnamefont {J.}~\bibnamefont {Mesot}}, \bibinfo {author} {\bibfnamefont {U.}~\bibnamefont {Staub}}, \bibinfo {author} {\bibfnamefont {M.}~\bibnamefont {Guillaume}}, \bibinfo {author} {\bibfnamefont {A.}~\bibnamefont {Furrer}}, \bibinfo {author} {\bibfnamefont {S.~I.}\ \bibnamefont {Yoo}}, \bibinfo {author} {\bibfnamefont {M.}~\bibnamefont {Kramer}}, \bibinfo {author} {\bibfnamefont {R.}~\bibnamefont {McCallum}}, \bibinfo {author} {\bibfnamefont {H.}~\bibnamefont {Maletta}}, \bibinfo {author} {\bibfnamefont {H.}~\bibnamefont {Blank}}, \emph {et~al.},\ }\bibfield  {title} {\bibinfo {title} {Magnetic properties of \ch{Nd^{3+}} in {N}d-{B}a-{C}u-{O}-compounds},\ }{\bibfield  {journal} {\bibinfo  {journal} {Zeitschrift f{\"u}r Physik B Condensed Matter}\ }\textbf {\bibinfo {volume} {95}},\ \bibinfo {pages} {301} (\bibinfo {year} {1994})}\BibitemShut {NoStop}%
\bibitem [{\citenamefont {Xiao}\ \emph {et~al.}(2012)\citenamefont {Xiao}, \citenamefont {Su}, \citenamefont {Nandi}, \citenamefont {Price}, \citenamefont {Schmitz}, \citenamefont {Kumar}, \citenamefont {Mittal}, \citenamefont {Chatterji}, \citenamefont {Kumar}, \citenamefont {Dhar} \emph {et~al.}}]{xiao2012anomalous}%
  \BibitemOpen
  \bibfield  {author} {\bibinfo {author} {\bibfnamefont {Y.}~\bibnamefont {Xiao}}, \bibinfo {author} {\bibfnamefont {Y.}~\bibnamefont {Su}}, \bibinfo {author} {\bibfnamefont {S.}~\bibnamefont {Nandi}}, \bibinfo {author} {\bibfnamefont {S.}~\bibnamefont {Price}}, \bibinfo {author} {\bibfnamefont {B.}~\bibnamefont {Schmitz}}, \bibinfo {author} {\bibfnamefont {C.}~\bibnamefont {Kumar}}, \bibinfo {author} {\bibfnamefont {R.}~\bibnamefont {Mittal}}, \bibinfo {author} {\bibfnamefont {T.}~\bibnamefont {Chatterji}}, \bibinfo {author} {\bibfnamefont {N.}~\bibnamefont {Kumar}}, \bibinfo {author} {\bibfnamefont {S.}~\bibnamefont {Dhar}}, \emph {et~al.},\ }\bibfield  {title} {\bibinfo {title} {Anomalous in-plane magnetoresistance in a \ch{EuFe2As2} single crystal: {E}vidence of strong spin-charge-lattice coupling},\ }{\bibfield  {journal} {\bibinfo  {journal} {Physical Review B}\ }\textbf {\bibinfo {volume} {85}},\ \bibinfo {pages} {094504} (\bibinfo {year}
  {2012})}\BibitemShut {NoStop}%
\bibitem [{\citenamefont {Hossain}\ \emph {et~al.}(2000)\citenamefont {Hossain}, \citenamefont {Hamashima}, \citenamefont {Umeo}, \citenamefont {Takabatake}, \citenamefont {Geibel},\ and\ \citenamefont {Steglich}}]{hossain2000antiferromagnetic}%
  \BibitemOpen
  \bibfield  {author} {\bibinfo {author} {\bibfnamefont {Z.}~\bibnamefont {Hossain}}, \bibinfo {author} {\bibfnamefont {S.}~\bibnamefont {Hamashima}}, \bibinfo {author} {\bibfnamefont {K.}~\bibnamefont {Umeo}}, \bibinfo {author} {\bibfnamefont {T.}~\bibnamefont {Takabatake}}, \bibinfo {author} {\bibfnamefont {C.}~\bibnamefont {Geibel}},\ and\ \bibinfo {author} {\bibfnamefont {F.}~\bibnamefont {Steglich}},\ }\bibfield  {title} {\bibinfo {title} {Antiferromagnetic transitions in the {K}ondo lattice system \ch{Ce2Ni3Ge5}},\ }{\bibfield  {journal} {\bibinfo  {journal} {Physical Review B}\ }\textbf {\bibinfo {volume} {62}},\ \bibinfo {pages} {8950} (\bibinfo {year} {2000})}\BibitemShut {NoStop}%
\bibitem [{\citenamefont {Shekhar}\ \emph {et~al.}(2018)\citenamefont {Shekhar}, \citenamefont {Kumar}, \citenamefont {Grinenko}, \citenamefont {Singh}, \citenamefont {Sarkar}, \citenamefont {Luetkens}, \citenamefont {Wu}, \citenamefont {Zhang}, \citenamefont {Komarek}, \citenamefont {Kampert} \emph {et~al.}}]{shekhar2018anomalous}%
  \BibitemOpen
  \bibfield  {author} {\bibinfo {author} {\bibfnamefont {C.}~\bibnamefont {Shekhar}}, \bibinfo {author} {\bibfnamefont {N.}~\bibnamefont {Kumar}}, \bibinfo {author} {\bibfnamefont {V.}~\bibnamefont {Grinenko}}, \bibinfo {author} {\bibfnamefont {S.}~\bibnamefont {Singh}}, \bibinfo {author} {\bibfnamefont {R.}~\bibnamefont {Sarkar}}, \bibinfo {author} {\bibfnamefont {H.}~\bibnamefont {Luetkens}}, \bibinfo {author} {\bibfnamefont {S.~C.}\ \bibnamefont {Wu}}, \bibinfo {author} {\bibfnamefont {Y.}~\bibnamefont {Zhang}}, \bibinfo {author} {\bibfnamefont {A.~C.}\ \bibnamefont {Komarek}}, \bibinfo {author} {\bibfnamefont {E.}~\bibnamefont {Kampert}}, \emph {et~al.},\ }\bibfield  {title} {\bibinfo {title} {Anomalous {H}all effect in Weyl semimetal half-heusler compounds \ch{RPtBi} ({R} = {G}d and {N}d)},\ }{\bibfield  {journal} {\bibinfo  {journal} {Proceedings of the National Academy of Sciences}\ }\textbf {\bibinfo {volume} {115}},\ \bibinfo {pages} {9140} (\bibinfo {year} {2018})}\BibitemShut
  {NoStop}%
\bibitem [{\citenamefont {Kotegawa}\ \emph {et~al.}(2023)\citenamefont {Kotegawa}, \citenamefont {Kuwata}, \citenamefont {Huyen}, \citenamefont {Arai}, \citenamefont {Tou}, \citenamefont {Matsuda}, \citenamefont {Takeda}, \citenamefont {Sugawara},\ and\ \citenamefont {Suzuki}}]{kotegawa2023large}%
  \BibitemOpen
  \bibfield  {author} {\bibinfo {author} {\bibfnamefont {H.}~\bibnamefont {Kotegawa}}, \bibinfo {author} {\bibfnamefont {Y.}~\bibnamefont {Kuwata}}, \bibinfo {author} {\bibfnamefont {V.~T.~N.}\ \bibnamefont {Huyen}}, \bibinfo {author} {\bibfnamefont {Y.}~\bibnamefont {Arai}}, \bibinfo {author} {\bibfnamefont {H.}~\bibnamefont {Tou}}, \bibinfo {author} {\bibfnamefont {M.}~\bibnamefont {Matsuda}}, \bibinfo {author} {\bibfnamefont {K.}~\bibnamefont {Takeda}}, \bibinfo {author} {\bibfnamefont {H.}~\bibnamefont {Sugawara}},\ and\ \bibinfo {author} {\bibfnamefont {M.-T.}\ \bibnamefont {Suzuki}},\ }\bibfield  {title} {\bibinfo {title} {Large anomalous {H}all effect and unusual domain switching in an orthorhombic antiferromagnetic material \ch{NbMnP}},\ }{\bibfield  {journal} {\bibinfo  {journal} {npj Quantum Materials}\ }\textbf {\bibinfo {volume} {8}},\ \bibinfo {pages} {56} (\bibinfo {year} {2023})}\BibitemShut {NoStop}%
\bibitem [{\citenamefont {Kim}\ \emph {et~al.}(2018)\citenamefont {Kim}, \citenamefont {Seo}, \citenamefont {Lee}, \citenamefont {Ko}, \citenamefont {Kim}, \citenamefont {Jang}, \citenamefont {Ok}, \citenamefont {Lee}, \citenamefont {Jo}, \citenamefont {Kang} \emph {et~al.}}]{kim2018large}%
  \BibitemOpen
  \bibfield  {author} {\bibinfo {author} {\bibfnamefont {K.}~\bibnamefont {Kim}}, \bibinfo {author} {\bibfnamefont {J.}~\bibnamefont {Seo}}, \bibinfo {author} {\bibfnamefont {E.}~\bibnamefont {Lee}}, \bibinfo {author} {\bibfnamefont {K.-T.}\ \bibnamefont {Ko}}, \bibinfo {author} {\bibfnamefont {B.}~\bibnamefont {Kim}}, \bibinfo {author} {\bibfnamefont {B.~G.}\ \bibnamefont {Jang}}, \bibinfo {author} {\bibfnamefont {J.~M.}\ \bibnamefont {Ok}}, \bibinfo {author} {\bibfnamefont {J.}~\bibnamefont {Lee}}, \bibinfo {author} {\bibfnamefont {Y.~J.}\ \bibnamefont {Jo}}, \bibinfo {author} {\bibfnamefont {W.}~\bibnamefont {Kang}}, \emph {et~al.},\ }\bibfield  {title} {\bibinfo {title} {Large anomalous {H}all current induced by topological nodal lines in a ferromagnetic Van der Waals semimetal},\ }{\bibfield  {journal} {\bibinfo  {journal} {Nature Materials}\ }\textbf {\bibinfo {volume} {17}},\ \bibinfo {pages} {794} (\bibinfo {year} {2018})}\BibitemShut {NoStop}%
\bibitem [{\citenamefont {Chatterjee}\ \emph {et~al.}(2023)\citenamefont {Chatterjee}, \citenamefont {Sau}, \citenamefont {Samanta}, \citenamefont {Ghosh}, \citenamefont {Kumar}, \citenamefont {Kumar},\ and\ \citenamefont {Mandal}}]{chatterjee2023nodal}%
  \BibitemOpen
  \bibfield  {author} {\bibinfo {author} {\bibfnamefont {S.}~\bibnamefont {Chatterjee}}, \bibinfo {author} {\bibfnamefont {J.}~\bibnamefont {Sau}}, \bibinfo {author} {\bibfnamefont {S.}~\bibnamefont {Samanta}}, \bibinfo {author} {\bibfnamefont {B.}~\bibnamefont {Ghosh}}, \bibinfo {author} {\bibfnamefont {N.}~\bibnamefont {Kumar}}, \bibinfo {author} {\bibfnamefont {M.}~\bibnamefont {Kumar}},\ and\ \bibinfo {author} {\bibfnamefont {K.}~\bibnamefont {Mandal}},\ }\bibfield  {title} {\bibinfo {title} {Nodal-line and triple point fermion induced anomalous {H}all effect in the topological {H}eusler compound \ch{Co2CrGa}},\ }{\bibfield  {journal} {\bibinfo  {journal} {Physical Review B}\ }\textbf {\bibinfo {volume} {107}},\ \bibinfo {pages} {125138} (\bibinfo {year} {2023})}\BibitemShut {NoStop}%
\bibitem [{\citenamefont {Karplus}\ and\ \citenamefont {Luttinger}(1954)}]{karplus1954hall}%
  \BibitemOpen
  \bibfield  {author} {\bibinfo {author} {\bibfnamefont {R.}~\bibnamefont {Karplus}}\ and\ \bibinfo {author} {\bibfnamefont {J.}~\bibnamefont {Luttinger}},\ }\bibfield  {title} {\bibinfo {title} {Hall effect in ferromagnetics},\ }{\bibfield  {journal} {\bibinfo  {journal} {Physical Review}\ }\textbf {\bibinfo {volume} {95}},\ \bibinfo {pages} {1154} (\bibinfo {year} {1954})}\BibitemShut {NoStop}%
\bibitem [{\citenamefont {Jungwirth}\ \emph {et~al.}(2002)\citenamefont {Jungwirth}, \citenamefont {Niu},\ and\ \citenamefont {MacDonald}}]{jungwirth2002anomalous}%
  \BibitemOpen
  \bibfield  {author} {\bibinfo {author} {\bibfnamefont {T.}~\bibnamefont {Jungwirth}}, \bibinfo {author} {\bibfnamefont {Q.}~\bibnamefont {Niu}},\ and\ \bibinfo {author} {\bibfnamefont {A.}~\bibnamefont {MacDonald}},\ }\bibfield  {title} {\bibinfo {title} {Anomalous {H}all effect in ferromagnetic semiconductors},\ }{\bibfield  {journal} {\bibinfo  {journal} {Physical Review Letters}\ }\textbf {\bibinfo {volume} {88}},\ \bibinfo {pages} {207208} (\bibinfo {year} {2002})}\BibitemShut {NoStop}%
\bibitem [{\citenamefont {Tian}\ \emph {et~al.}(2009)\citenamefont {Tian}, \citenamefont {Ye},\ and\ \citenamefont {Jin}}]{tian2009proper}%
  \BibitemOpen
  \bibfield  {author} {\bibinfo {author} {\bibfnamefont {Y.}~\bibnamefont {Tian}}, \bibinfo {author} {\bibfnamefont {L.}~\bibnamefont {Ye}},\ and\ \bibinfo {author} {\bibfnamefont {X.}~\bibnamefont {Jin}},\ }\bibfield  {title} {\bibinfo {title} {Proper scaling of the anomalous {H}all effect},\ } {\bibfield  {journal} {\bibinfo  {journal} {Physical Review Letters}\ }\textbf {\bibinfo {volume} {103}},\ \bibinfo {pages} {087206} (\bibinfo {year} {2009})}\BibitemShut {NoStop}%
\bibitem [{\citenamefont {Hou}\ \emph {et~al.}(2015)\citenamefont {Hou}, \citenamefont {Su}, \citenamefont {Tian}, \citenamefont {Jin}, \citenamefont {Yang},\ and\ \citenamefont {Niu}}]{hou2015multivariable}%
  \BibitemOpen
  \bibfield  {author} {\bibinfo {author} {\bibfnamefont {D.}~\bibnamefont {Hou}}, \bibinfo {author} {\bibfnamefont {G.}~\bibnamefont {Su}}, \bibinfo {author} {\bibfnamefont {Y.}~\bibnamefont {Tian}}, \bibinfo {author} {\bibfnamefont {X.}~\bibnamefont {Jin}}, \bibinfo {author} {\bibfnamefont {S.~A.}\ \bibnamefont {Yang}},\ and\ \bibinfo {author} {\bibfnamefont {Q.}~\bibnamefont {Niu}},\ }\bibfield  {title} {\bibinfo {title} {Multivariable scaling for the anomalous {H}all effect},\ }{\bibfield  {journal} {\bibinfo  {journal} {Physical Review Letters}\ }\textbf {\bibinfo {volume} {114}},\ \bibinfo {pages} {217203} (\bibinfo {year} {2015})}\BibitemShut {NoStop}%
\bibitem [{\citenamefont {Yang}\ \emph {et~al.}(2020)\citenamefont {Yang}, \citenamefont {Singh}, \citenamefont {Lu}, \citenamefont {Huang}, \citenamefont {Bahrami}, \citenamefont {Chiu}, \citenamefont {Graf}, \citenamefont {Huang}, \citenamefont {Wang}, \citenamefont {Lin} \emph {et~al.}}]{yang2020transition}%
  \BibitemOpen
  \bibfield  {author} {\bibinfo {author} {\bibfnamefont {H.-Y.}\ \bibnamefont {Yang}}, \bibinfo {author} {\bibfnamefont {B.}~\bibnamefont {Singh}}, \bibinfo {author} {\bibfnamefont {B.}~\bibnamefont {Lu}}, \bibinfo {author} {\bibfnamefont {C.-Y.}\ \bibnamefont {Huang}}, \bibinfo {author} {\bibfnamefont {F.}~\bibnamefont {Bahrami}}, \bibinfo {author} {\bibfnamefont {W.-C.}\ \bibnamefont {Chiu}}, \bibinfo {author} {\bibfnamefont {D.}~\bibnamefont {Graf}}, \bibinfo {author} {\bibfnamefont {S.-M.}\ \bibnamefont {Huang}}, \bibinfo {author} {\bibfnamefont {B.}~\bibnamefont {Wang}}, \bibinfo {author} {\bibfnamefont {H.}~\bibnamefont {Lin}}, \emph {et~al.},\ }\bibfield  {title} {\bibinfo {title} {Transition from intrinsic to extrinsic anomalous {H}all effect in the ferromagnetic {W}eyl semimetal PrAlGe$_{1-x}$Si$_x$},\ }{\bibfield  {journal} {\bibinfo  {journal} {APL Materials}\ }\textbf {\bibinfo {volume} {8}} (\bibinfo {year} {2020})}\BibitemShut {NoStop}%
\bibitem [{\citenamefont {Zhou}\ \emph {et~al.}(2023)\citenamefont {Zhou}, \citenamefont {Shi}, \citenamefont {Huang}, \citenamefont {Ma}, \citenamefont {Xu}, \citenamefont {Wang},\ and\ \citenamefont {Jia}}]{zhou2023metamagnetic}%
  \BibitemOpen
  \bibfield  {author} {\bibinfo {author} {\bibfnamefont {H.}~\bibnamefont {Zhou}}, \bibinfo {author} {\bibfnamefont {M.}~\bibnamefont {Shi}}, \bibinfo {author} {\bibfnamefont {Y.}~\bibnamefont {Huang}}, \bibinfo {author} {\bibfnamefont {W.}~\bibnamefont {Ma}}, \bibinfo {author} {\bibfnamefont {X.}~\bibnamefont {Xu}}, \bibinfo {author} {\bibfnamefont {J.}~\bibnamefont {Wang}},\ and\ \bibinfo {author} {\bibfnamefont {S.}~\bibnamefont {Jia}},\ }\bibfield  {title} {\bibinfo {title} {Metamagnetic transition and anomalous {H}all effect in {M}n-based kagom{\'e} magnets \ch{RMn6Ge6} ({R} = {T}b-{L}u)},\ } {\bibfield  {journal} {\bibinfo  {journal} {Physical Review Materials}\ }\textbf {\bibinfo {volume} {7}},\ \bibinfo {pages} {024404} (\bibinfo {year} {2023})}\BibitemShut {NoStop}%
\bibitem [{\citenamefont {Liu}\ \emph {et~al.}(2018)\citenamefont {Liu}, \citenamefont {Sun}, \citenamefont {Kumar}, \citenamefont {Muechler}, \citenamefont {Sun}, \citenamefont {Jiao}, \citenamefont {Yang}, \citenamefont {Liu}, \citenamefont {Liang}, \citenamefont {Xu} \emph {et~al.}}]{liu2018giant}%
  \BibitemOpen
  \bibfield  {author} {\bibinfo {author} {\bibfnamefont {E.}~\bibnamefont {Liu}}, \bibinfo {author} {\bibfnamefont {Y.}~\bibnamefont {Sun}}, \bibinfo {author} {\bibfnamefont {N.}~\bibnamefont {Kumar}}, \bibinfo {author} {\bibfnamefont {L.}~\bibnamefont {Muechler}}, \bibinfo {author} {\bibfnamefont {A.}~\bibnamefont {Sun}}, \bibinfo {author} {\bibfnamefont {L.}~\bibnamefont {Jiao}}, \bibinfo {author} {\bibfnamefont {S.-Y.}\ \bibnamefont {Yang}}, \bibinfo {author} {\bibfnamefont {D.}~\bibnamefont {Liu}}, \bibinfo {author} {\bibfnamefont {A.}~\bibnamefont {Liang}}, \bibinfo {author} {\bibfnamefont {Q.}~\bibnamefont {Xu}}, \emph {et~al.},\ }\bibfield  {title} {\bibinfo {title} {Giant anomalous {H}all effect in a ferromagnetic kagom{\'e}-lattice semimetal},\ }{\bibfield  {journal} {\bibinfo  {journal} {Nature Physics}\ }\textbf {\bibinfo {volume} {14}},\ \bibinfo {pages} {1125} (\bibinfo {year} {2018})}\BibitemShut {NoStop}%
\bibitem [{\citenamefont {Ye}\ \emph {et~al.}(2018)\citenamefont {Ye}, \citenamefont {Kang}, \citenamefont {Liu}, \citenamefont {Von~Cube}, \citenamefont {Wicker}, \citenamefont {Suzuki}, \citenamefont {Jozwiak}, \citenamefont {Bostwick}, \citenamefont {Rotenberg}, \citenamefont {Bell} \emph {et~al.}}]{ye2018massive}%
  \BibitemOpen
  \bibfield  {author} {\bibinfo {author} {\bibfnamefont {L.}~\bibnamefont {Ye}}, \bibinfo {author} {\bibfnamefont {M.}~\bibnamefont {Kang}}, \bibinfo {author} {\bibfnamefont {J.}~\bibnamefont {Liu}}, \bibinfo {author} {\bibfnamefont {F.}~\bibnamefont {Von~Cube}}, \bibinfo {author} {\bibfnamefont {C.~R.}\ \bibnamefont {Wicker}}, \bibinfo {author} {\bibfnamefont {T.}~\bibnamefont {Suzuki}}, \bibinfo {author} {\bibfnamefont {C.}~\bibnamefont {Jozwiak}}, \bibinfo {author} {\bibfnamefont {A.}~\bibnamefont {Bostwick}}, \bibinfo {author} {\bibfnamefont {E.}~\bibnamefont {Rotenberg}}, \bibinfo {author} {\bibfnamefont {D.~C.}\ \bibnamefont {Bell}}, \emph {et~al.},\ }\bibfield  {title} {\bibinfo {title} {Massive {D}irac fermions in a ferromagnetic kagom{\'e} metal},\ } {\bibfield  {journal} {\bibinfo  {journal} {Nature}\ }\textbf {\bibinfo {volume} {555}},\ \bibinfo {pages} {638} (\bibinfo {year} {2018})}\BibitemShut {NoStop}%
\bibitem [{\citenamefont {Wang}\ \emph {et~al.}(2017)\citenamefont {Wang}, \citenamefont {Xian}, \citenamefont {Wang}, \citenamefont {Liu}, \citenamefont {Ling}, \citenamefont {Zhang}, \citenamefont {Cao}, \citenamefont {Qu},\ and\ \citenamefont {Xiong}}]{wang2017anisotropic}%
  \BibitemOpen
  \bibfield  {author} {\bibinfo {author} {\bibfnamefont {Y.}~\bibnamefont {Wang}}, \bibinfo {author} {\bibfnamefont {C.}~\bibnamefont {Xian}}, \bibinfo {author} {\bibfnamefont {J.}~\bibnamefont {Wang}}, \bibinfo {author} {\bibfnamefont {B.}~\bibnamefont {Liu}}, \bibinfo {author} {\bibfnamefont {L.}~\bibnamefont {Ling}}, \bibinfo {author} {\bibfnamefont {L.}~\bibnamefont {Zhang}}, \bibinfo {author} {\bibfnamefont {L.}~\bibnamefont {Cao}}, \bibinfo {author} {\bibfnamefont {Z.}~\bibnamefont {Qu}},\ and\ \bibinfo {author} {\bibfnamefont {Y.}~\bibnamefont {Xiong}},\ }\bibfield  {title} {\bibinfo {title} {Anisotropic anomalous {H}all effect in triangular itinerant ferromagnet \ch{Fe3GeTe2}},\ }{\bibfield  {journal} {\bibinfo  {journal} {Physical Review B}\ }\textbf {\bibinfo {volume} {96}},\ \bibinfo {pages} {134428} (\bibinfo {year} {2017})}\BibitemShut {NoStop}%
\bibitem [{\citenamefont {Li}\ \emph {et~al.}(2020)\citenamefont {Li}, \citenamefont {Zhang}, \citenamefont {Liang}, \citenamefont {Ding}, \citenamefont {Chen}, \citenamefont {Shen}, \citenamefont {Li}, \citenamefont {Liu}, \citenamefont {Xi}, \citenamefont {Wu} \emph {et~al.}}]{li2020large}%
  \BibitemOpen
  \bibfield  {author} {\bibinfo {author} {\bibfnamefont {H.}~\bibnamefont {Li}}, \bibinfo {author} {\bibfnamefont {B.}~\bibnamefont {Zhang}}, \bibinfo {author} {\bibfnamefont {J.}~\bibnamefont {Liang}}, \bibinfo {author} {\bibfnamefont {B.}~\bibnamefont {Ding}}, \bibinfo {author} {\bibfnamefont {J.}~\bibnamefont {Chen}}, \bibinfo {author} {\bibfnamefont {J.}~\bibnamefont {Shen}}, \bibinfo {author} {\bibfnamefont {Z.}~\bibnamefont {Li}}, \bibinfo {author} {\bibfnamefont {E.}~\bibnamefont {Liu}}, \bibinfo {author} {\bibfnamefont {X.}~\bibnamefont {Xi}}, \bibinfo {author} {\bibfnamefont {G.}~\bibnamefont {Wu}}, \emph {et~al.},\ }\bibfield  {title} {\bibinfo {title} {Large anomalous {H}all effect in a hexagonal ferromagnetic \ch{Fe5Sn3} single crystal},\ }{\bibfield  {journal} {\bibinfo  {journal} {Physical Review B}\ }\textbf {\bibinfo {volume} {101}},\ \bibinfo {pages} {140409} (\bibinfo {year} {2020})}\BibitemShut {NoStop}%
\bibitem [{\citenamefont {Bera}\ \emph {et~al.}(2023)\citenamefont {Bera}, \citenamefont {Chatterjee}, \citenamefont {Pradhan}, \citenamefont {Pradhan}, \citenamefont {Kalimuddin}, \citenamefont {Bera}, \citenamefont {Nandy},\ and\ \citenamefont {Mondal}}]{bera2023anomalous}%
  \BibitemOpen
  \bibfield  {author} {\bibinfo {author} {\bibfnamefont {S.}~\bibnamefont {Bera}}, \bibinfo {author} {\bibfnamefont {S.}~\bibnamefont {Chatterjee}}, \bibinfo {author} {\bibfnamefont {S.}~\bibnamefont {Pradhan}}, \bibinfo {author} {\bibfnamefont {S.~K.}\ \bibnamefont {Pradhan}}, \bibinfo {author} {\bibfnamefont {S.}~\bibnamefont {Kalimuddin}}, \bibinfo {author} {\bibfnamefont {A.}~\bibnamefont {Bera}}, \bibinfo {author} {\bibfnamefont {A.~K.}\ \bibnamefont {Nandy}},\ and\ \bibinfo {author} {\bibfnamefont {M.}~\bibnamefont {Mondal}},\ }\bibfield  {title} {\bibinfo {title} {Anomalous {H}all effect induced by {B}erry curvature in the topological nodal-line van der waals ferromagnet \ch{Fe4GeTe2}},\ }{\bibfield  {journal} {\bibinfo  {journal} {Physical Review B}\ }\textbf {\bibinfo {volume} {108}},\ \bibinfo {pages} {115122} (\bibinfo {year} {2023})}\BibitemShut {NoStop}%
\bibitem [{\citenamefont {Alam}\ \emph {et~al.}(2023)\citenamefont {Alam}, \citenamefont {Fakhredine}, \citenamefont {Ahmad}, \citenamefont {Tanwar}, \citenamefont {Yang}, \citenamefont {Tafti}, \citenamefont {Cuono}, \citenamefont {Islam}, \citenamefont {Singh}, \citenamefont {Lynnyk} \emph {et~al.}}]{alam2023sign}%
  \BibitemOpen
  \bibfield  {author} {\bibinfo {author} {\bibfnamefont {M.~S.}\ \bibnamefont {Alam}}, \bibinfo {author} {\bibfnamefont {A.}~\bibnamefont {Fakhredine}}, \bibinfo {author} {\bibfnamefont {M.}~\bibnamefont {Ahmad}}, \bibinfo {author} {\bibfnamefont {P.}~\bibnamefont {Tanwar}}, \bibinfo {author} {\bibfnamefont {H.~Y.}\ \bibnamefont {Yang}}, \bibinfo {author} {\bibfnamefont {F.}~\bibnamefont {Tafti}}, \bibinfo {author} {\bibfnamefont {G.}~\bibnamefont {Cuono}}, \bibinfo {author} {\bibfnamefont {R.}~\bibnamefont {Islam}}, \bibinfo {author} {\bibfnamefont {B.}~\bibnamefont {Singh}}, \bibinfo {author} {\bibfnamefont {A.}~\bibnamefont {Lynnyk}}, \emph {et~al.},\ }\bibfield  {title} {\bibinfo {title} {Sign change of anomalous {H}all effect and anomalous {N}ernst effect in the {W}eyl semimetal \ch{CeAlSi}},\ }{\bibfield  {journal} {\bibinfo  {journal} {Physical Review B}\ }\textbf {\bibinfo {volume} {107}},\ \bibinfo {pages} {085102} (\bibinfo {year} {2023})}\BibitemShut {NoStop}%
\bibitem [{\citenamefont {Wang}\ \emph {et~al.}(2021)\citenamefont {Wang}, \citenamefont {Neubauer}, \citenamefont {Duan}, \citenamefont {Yin}, \citenamefont {Fujitsu}, \citenamefont {Hosono}, \citenamefont {Ye}, \citenamefont {Zhang}, \citenamefont {Chi}, \citenamefont {Krycka} \emph {et~al.}}]{wang2021field}%
  \BibitemOpen
  \bibfield  {author} {\bibinfo {author} {\bibfnamefont {Q.}~\bibnamefont {Wang}}, \bibinfo {author} {\bibfnamefont {K.~J.}\ \bibnamefont {Neubauer}}, \bibinfo {author} {\bibfnamefont {C.}~\bibnamefont {Duan}}, \bibinfo {author} {\bibfnamefont {Q.}~\bibnamefont {Yin}}, \bibinfo {author} {\bibfnamefont {S.}~\bibnamefont {Fujitsu}}, \bibinfo {author} {\bibfnamefont {H.}~\bibnamefont {Hosono}}, \bibinfo {author} {\bibfnamefont {F.}~\bibnamefont {Ye}}, \bibinfo {author} {\bibfnamefont {R.}~\bibnamefont {Zhang}}, \bibinfo {author} {\bibfnamefont {S.}~\bibnamefont {Chi}}, \bibinfo {author} {\bibfnamefont {K.}~\bibnamefont {Krycka}}, \emph {et~al.},\ }\bibfield  {title} {\bibinfo {title} {Field-induced topological Hall effect and double-fan spin structure with ac-axis component in the metallic kagom{\'e} antiferromagnetic compound \ch{YMn6Sn6}},\ }{\bibfield  {journal} {\bibinfo  {journal} {Physical Review B}\ }\textbf {\bibinfo {volume} {103}},\ \bibinfo {pages} {014416} (\bibinfo {year}
  {2021})}\BibitemShut {NoStop}%
\bibitem [{\citenamefont {Arai}\ \emph {et~al.}(2024)\citenamefont {Arai}, \citenamefont {Hayashi}, \citenamefont {Takeda}, \citenamefont {Tou}, \citenamefont {Sugawara},\ and\ \citenamefont {Kotegawa}}]{arai2024intrinsic}%
  \BibitemOpen
  \bibfield  {author} {\bibinfo {author} {\bibfnamefont {Y.}~\bibnamefont {Arai}}, \bibinfo {author} {\bibfnamefont {J.}~\bibnamefont {Hayashi}}, \bibinfo {author} {\bibfnamefont {K.}~\bibnamefont {Takeda}}, \bibinfo {author} {\bibfnamefont {H.}~\bibnamefont {Tou}}, \bibinfo {author} {\bibfnamefont {H.}~\bibnamefont {Sugawara}},\ and\ \bibinfo {author} {\bibfnamefont {H.}~\bibnamefont {Kotegawa}},\ }\bibfield  {title} {\bibinfo {title} {Intrinsic anomalous {H}all {E}ffect arising from antiferromagnetism as revealed by high-quality \ch{NbMnP}},\ }{\bibfield  {journal} {\bibinfo  {journal} {Journal of the Physical Society of Japan}\ }\textbf {\bibinfo {volume} {93}},\ \bibinfo {pages} {063702} (\bibinfo {year} {2024})}\BibitemShut {NoStop}%
\bibitem [{\citenamefont {Suzuki}\ \emph {et~al.}(2016)\citenamefont {Suzuki}, \citenamefont {Chisnell}, \citenamefont {Devarakonda}, \citenamefont {Liu}, \citenamefont {Feng}, \citenamefont {Xiao}, \citenamefont {Lynn},\ and\ \citenamefont {Checkelsky}}]{suzuki2016large}%
  \BibitemOpen
  \bibfield  {author} {\bibinfo {author} {\bibfnamefont {T.}~\bibnamefont {Suzuki}}, \bibinfo {author} {\bibfnamefont {R.}~\bibnamefont {Chisnell}}, \bibinfo {author} {\bibfnamefont {A.}~\bibnamefont {Devarakonda}}, \bibinfo {author} {\bibfnamefont {Y.-T.}\ \bibnamefont {Liu}}, \bibinfo {author} {\bibfnamefont {W.}~\bibnamefont {Feng}}, \bibinfo {author} {\bibfnamefont {D.}~\bibnamefont {Xiao}}, \bibinfo {author} {\bibfnamefont {J.~W.}\ \bibnamefont {Lynn}},\ and\ \bibinfo {author} {\bibfnamefont {J.}~\bibnamefont {Checkelsky}},\ }\bibfield  {title} {\bibinfo {title} {Large anomalous {H}all effect in a half-Heusler antiferromagnet},\ }{\bibfield  {journal} {\bibinfo  {journal} {Nature Physics}\ }\textbf {\bibinfo {volume} {12}},\ \bibinfo {pages} {1119} (\bibinfo {year} {2016})}\BibitemShut {NoStop}%
\bibitem [{\citenamefont {Zeng}\ \emph {et~al.}(2022)\citenamefont {Zeng}, \citenamefont {Yu}, \citenamefont {Luo}, \citenamefont {Chen}, \citenamefont {Fang}, \citenamefont {Ma}, \citenamefont {Mo}, \citenamefont {Shen}, \citenamefont {Yuan},\ and\ \citenamefont {Zhong}}]{zeng2022large}%
  \BibitemOpen
  \bibfield  {author} {\bibinfo {author} {\bibfnamefont {H.}~\bibnamefont {Zeng}}, \bibinfo {author} {\bibfnamefont {G.}~\bibnamefont {Yu}}, \bibinfo {author} {\bibfnamefont {X.}~\bibnamefont {Luo}}, \bibinfo {author} {\bibfnamefont {C.}~\bibnamefont {Chen}}, \bibinfo {author} {\bibfnamefont {C.}~\bibnamefont {Fang}}, \bibinfo {author} {\bibfnamefont {S.}~\bibnamefont {Ma}}, \bibinfo {author} {\bibfnamefont {Z.}~\bibnamefont {Mo}}, \bibinfo {author} {\bibfnamefont {J.}~\bibnamefont {Shen}}, \bibinfo {author} {\bibfnamefont {M.}~\bibnamefont {Yuan}},\ and\ \bibinfo {author} {\bibfnamefont {Z.}~\bibnamefont {Zhong}},\ }\bibfield  {title} {\bibinfo {title} {Large anomalous {H}all effect in kagom{\'e} ferrimagnetic \ch{HoMn6Sn6} single crystal},\ } {\bibfield  {journal} {\bibinfo  {journal} {Journal of Alloys and Compounds}\ }\textbf {\bibinfo {volume} {899}},\ \bibinfo {pages} {163356} (\bibinfo {year} {2022})}\BibitemShut {NoStop}%
\end{thebibliography}
\end{document}